%% file: main.tex
\documentclass[copyright,creativecommons]{eptcs}

\usepackage{color}
\usepackage{graphicx}
\usepackage{calc,amsmath,latexsym,amssymb}
\usepackage{amsthm}
\usepackage{stmaryrd}
\usepackage{tabularx}
\usepackage{algorithm}
\usepackage{algorithmic}
\input{macros}

\title{Implementing Distributed Controllers for Systems with Priorities}

\author{Imene Ben-Hafaiedh \qquad\qquad  Susanne Graf
\institute{Universite Joseph Fourier/VERIMAG \\ Grenoble, France}
\email{benhfaie@imag.fr \quad\qquad graf@imag.fr}
\and
  Hammadi Khairallah
\institute{Tunisia Polytechnic School\\
Tunis, Tunisia}
\email{\quad khairall@imag.fr}
}

\def\titlerunning{Implementing Distributed Controllers for Systems with Priorities}
\def\authorrunning{I. Ben-Hafaiedh, S. Graf, H. Khairallah}

\begin{document}

\maketitle

\input{abstract}


\section{Introduction}

\input{intro}

\section{Distributed Controllers for Systems with Priorities}
 \label{sec:controller}
 \input{preliminaries}

 \input{description}

\vspace{-0.5\baselineskip}
\section{The protocol}
\label{sec: protocol}
 \input{BIPImplementation}
 \input{cycles}

\vspace{-0.5\baselineskip}
\section{Implementation and experimental results}
\label{sec: expriments}

\input{implementation}

\input{degreeConflict}

\input{diningPhilo}


\vspace{-0.5\baselineskip}
\section{Conclusion and future work}
\label{sec:methodo}

\input{conclusion}


\bibliographystyle{eptcs}
\bibliography{bib}


 \input{algorithms}

\end{document}

%% file: macros.tex
\definecolor{tblue}{rgb}{0,0,.7}

\definecolor{lred}{rgb}{0.2,0,0}

\newcommand{\comment}[1]{}

\newcommand{\ignore}[1]{} 

\newcommand{\Ports}{\mathcal{P}}

\newtheorem{definition}{Definition}

\newcommand{\COMMIT}{\mathit{COMMIT }}
\newcommand{\REFUSE}{\mathit{REFUSE }}

\newcommand{\POSSIBLE}{\mathit{POSSIBLE }}
\newcommand{\NOTPOSSIBLE}{\mathit{NOTPOSSIBLE }}

\newcommand{\READY}{\mathit{READY }}
\newcommand{\NOTREADY}{\mathit{NOTREADY }}

%% file: abstract.tex
\begin{abstract}

Implementing a component-based system in a distributed way so that 
it ensures some global constraints is a challenging problem.
We consider here abstract specifications consisting of 
a composition of components and 
a controller given in the form of a set of interactions and a priority 
order amongst them. 
In the context of distributed systems, such a controller  must 
be executed in a distributed fashion while still respecting the global 
constraints imposed by interactions and priorities.
 We present in this paper an implementation of an algorithm
 that allows a distributed execution of systems
with (binary) interactions and priorities. We also present a comprehensive simulation analysis that shows how sensitive 
to changes our algorithm is, in particular changes related to the degree of conflict in the system.

\end{abstract}

%% file: intro.tex
A distributed system is a collection of 
components, or processes,
 communicating by explicit message passing. These components are intrinsically concurrent 
and knowledge about their respective states can be obtained only through communication. 
Thus determining the exact global state 
of such systems is not a trivial task \cite{ChandyLamport85}, and in general not required. 
The motivation of this work is to generate a distributed implementation of
 systems defined as a set of synchronizing processes and a set of
{\em priority} constraints among these synchronizations or interactions.

Specifying priorities amongst a set of alternative system interactions is interesting
in different contexts. 
For example, it is likely that amongst a set of enabled 
synchronizations amongst subsets of components, one will prefer those involving
larger subsets. Another typical example of the use of priorities are
 processes which for different activities require
one or more resources amongst a shared pool of resources. 

There exist several abstract frameworks allowing to represent specifications with 
priorities, such as process algebras with priorities or prioritized Petrinets. We present here
our results for a simple formalism, motivated by the application to BIP \cite{GoesslerS05,BasuBS06}.
In \cite{TRJLAP}, we have proposed an algorithm defining a controller for each of the processes and
that allows a distributed execution of such prioritized specification.
 
Here, we discuss an implementation of this algorithm and an evaluation of different  performance metrics. 
In a number of experiments, we measure the {\em execution-time}
 and {\em message-count} of the algorithm as a result of variations in different model 
parameters, in particular, variations in the degree of {\em conflict} of the system. We also
compare our algorithm to
 $\alpha$-core algorithm based on experimental results provided in \cite{PerezCT04}.

The paper is organized as follows: Section~\ref{sec:controller} introduces
the specification formalism and the relevant notions of concurrency and conflict for discussing 
the correctness of a distributed implementation, and then a description of the problem tackled by our algorithm. 
Section~\ref{sec: protocol} 
gives a brief description of the algorithm and how it works on an illustrative example.
 
The main contribution of this paper
is given in Section~\ref{sec: expriments} which describes the implementation of the algorithm 
and our experimental results.

%% file: preliminaries.tex
 \subsection{Processes interacting through synchronizations and priorities}

As a formalism for describing systems with synchronizations and priorities --- which we then want to 
execute in a distributed fashion --- we choose a very basic formalism for describing systems 
consisting of a set of components or processes which interact through n-ary synchronizations, 
and where there may be given a partial order on the set of synchronizations defining a priority 
amongst them. The motivation for this work roots in the need for algorithms for distributed executions of
 BIP models \cite{GoesslerS05,BasuBS06,BliudzeS07} or for prioritized Petrinets. Indeed, there are
straightforward mappings between the simple formalism considered here and the before mentioned
 more evolved ones. We consider processes to be represented by labeled transition systems where labels
represent a set of interactions on which several processes synchronize. That is, our formalism is close 
to Petrinets where interactions correspond to (joint) transitions.  On the other hand, in BIP we would 
associate  globally unique port names with components and identify an interaction by a set of ports of
different components. In terms of reusability BIP components are preferable, but here our aim is to 
be able to define the problem to be solved and the algorithm solving it in a most intuitive manner. And for this
purpose, naming interactions is most appropriate.
\begin{definition}[Process]
\label{def: atomic}
A process is a {\em Labeled Transition System} (LTS) represented by
a tuple $(Q,q^0,\Ports,\delta)$ where $Q$ is a set of states, $q^0\in Q$ is an
initial state, $\Ports$ is a set of labels representing interactions
and $\delta \subseteq Q\times \Ports\times Q$ is a transition relation.
\end{definition}
\vspace{-0.5\baselineskip}
As usually, we write $q_1\stackrel{a}{\longrightarrow} q_2$ instead of  $(q_1,a,q_2)\in \delta$ and
$q_1\stackrel{a}{\longrightarrow}$ instead of $\exists q'\in Q,
q\stackrel{a}{\longrightarrow} q'$. We also write sometimes $q_1 \stackrel{\epsilon}{\longrightarrow} q_2$
when $q_2 = q_1$.
\vspace{-\baselineskip}
\paragraph{Composed systems} Given a set of $n$ processes
 $K_i=(Q_i,q^0_i,\Ports_i,\delta_i)$ for $i\in [1,n]$, 
its composition is a process defined on the set of interactions  
$\Ports=\bigcup^n_{i=1}\Ports_i$\footnote{in fact, we choose a strict subset which means that
decide to block a set of interactions offered by some of the components and we can rename
some interactions to $\tau$ which means making them local; but this is not important for our algorithm.}.

 \begin{definition}[Interleaving semantics of composition]
The composition of $n$ processes $K_i$ 

 is denoted $K=\|(K_1,...\, ,K_n)$ and its semantics is defined by the  LTS $(Q, q^0,
  \Ports, \delta)$ where $Q = \prod_{i=1}^{n}Q_i$, $q^0 = (q^0_1,
  ...\,, q^0_n)$, $\Ports=\bigcup_{i=1}^n \Ports_i$. 
Now the transition relation  $\delta\subseteq Q\times\Ports\times Q$ is defined for each $a\in\Ports$ as
 the smallest subset of transitions  obtained by applying the following rule, where $I$ is the set of indexes of
the processes having $a$ in their alphabet:
 \begin{displaymath}
  \frac{\forall i\in I.\ q^1_i \stackrel{a}{\longrightarrow} q^2_i \wedge \ 
     \forall i\not \in I.\ q^1_i=q^2_i}
   {(q^1_1, ...\,, q^1_n) \stackrel{a}{\longrightarrow}(q^2_1, ...\,, q^2_n)}
   \end{displaymath}
\end{definition}

This means that a transition from state $q$ in the composed system  that $\|(K_1,...\, ,K_n)$
consists for any   $a\in \Ports$ of the joint execution of an $a$-transition in all the processes $K_i$ having 
$a$ in their alphabet. That is the fact that an $a$-interaction can be fired is defined locally in the
substate of the processes involved in this interaction.
\vspace{-0.5\baselineskip}
\paragraph{Priorities} A (partial) priority order amongst interaction may restrict the choice
of interactions that can be fired in a given state. That is, priorities may restrict 
nondeterminism. The semantics of priorities is global, that is, in presence of priority, whether a 
transition is enabled may depend on the entire global state.
\begin{definition}[Priority order]
\label{def:prio}
A {\em priority order} denoted by $<$ is a strict partial order on a set of interactions.
We denote that an interaction
$a$ has lower priority than $b$ by $a<b$.
\end{definition}

 \begin{definition}[System controlled by a priority order]
The semantics of a system $S=(Q, q^0,\Ports, \delta)$ {\em controlled} by a priority order $<$ 
defines an LTS $(Q, q^0,\Ports, \delta_<)$ 
   where $\delta_<$ is defined by the following rule, where we denote by $I$ the indexes of the 
processes involved in $a$ and by $q_I$ the substate of $q$ defined by $I$:
   \begin{displaymath}
     \frac{q_{I} \stackrel{a}{\longrightarrow}_{S} q'_{I}\ \wedge \  \nexists b\in \Ports.\ (a<b\wedge \  q\stackrel{b}{\longrightarrow})}{q \stackrel{a}{\longrightarrow_{<}} q'}
   \end{displaymath}
\end{definition}
Thus, only interactions that are locally enabled in all
concerned components, and furthermore not inhibited by an
interaction with higher priority, may be fired. An interesting property of priorities is the well-known fact
that they allow restricting the behavior of a system by guaranteeing that no new deadlocks are introduced 
by this restriction. This is the reason why we want to use priorities to ``control'' systems.
We denote the resulting controlled system by $(S,<)$. \medskip

We now introduce notations allowing to distinguish between the enabledness of a transition
locally in some process, in the uncontrolled system $S$ and in the controlled system $(S,<)$, where we suppose in the following $S$ to be defined as the composition of $P_i=(Q_i,q^0_i,\Ports_i,\delta_i)$ with $i\in [1,n]$
and $\Ports=\cup\Ports_i$.

\begin{definition}[locally ready, globally ready, enabled interaction]
\label{enabled}
Consider a global state $q\in Q_S$
such that $q=(q_1,...\,,q_n)$ and an interaction $a\in \Ports$.
\begin{itemize}
\item For $i$ such that $a\in \Ports_i$, $a$ is {\em locally ready} in $q_i$ iff 
  $\exists \ q_i'\in Q_i,\  s.t.\ q_i \stackrel{a}{\longrightarrow}_{i} q_i'$
\item $a$ is {\em  globally ready} in $q$ iff $\exists  q'\in Q_S.\ q \stackrel{a}{\longrightarrow}_{S}
    q'$ 
\item $a$ is {\em  enabled} in $q$ iff $a$ is {\em  globally ready} in $q$
and no interaction with higher priority is also globally ready in $q$, that is, iff
 $q\stackrel{a}{\longrightarrow}_{(S,<)}$.
\end{itemize}
\end{definition}

Note that only enabledness is related to priorities, and enabledness of $a$ implies global readiness which in turn
implies local readiness in all processes which have $a$ in their alphabet.
We are interested in distributed executions of $(S,<)$. We therefore define a notion of {\em concurrency} 
and {\em conflict} of interactions, such that in a distributed setting
we may allow the independent execution of concurrent interactions (so as to avoid global sequencing). 
We distinguish
explicitly between the usual notion of conflict which we call structural conflict, and a conflict due to
priorities.

\begin{definition}[Concurrent and conflicting  interactions]
\label{concurrent} \label{conflict} \label{prioconflict}
Let $a,b$ be  interactions of $\Ports$ and $q\in Q$ a global state in which $a$ and $b$ are globally ready. 
\begin{itemize}
\item $a$ and $b$ are called {\em concurrent} in  $q$ iff $a$ and $b$ are globally ready in $q$
and the set of processes $I_a$, $I_b$ involved in $a$, resp. $b$ are disjoint.

That is, when $a$ is executed then $b$ is still globally ready
afterwards, and vice versa, and if executed, both execution sequences lead to the same global state.

\item $a$ and $b$ are called {\em in structural conflict} in  $q$ iff
they are not concurrent in $q$, that is $a$ and $b$ are alternatives disabling each other.

\item $a$ and $b$ are  in {\em prioritized conflict} in $q$ iff $a$ and $b$ are concurrent in $q$
      but $a<b$ or $b<a$ holds.
\end{itemize}

\end{definition}
Note that in case of prioritized conflict, it is known which interaction
cannot be executed, whereas in case of structural conflict, the situation is symmetric.
Note that there is a particular situation, called a prioritized confusion, in which $a$ and $b$ are concurrent
 and both of maximal priority in $q$, but when $a$ is executed a state $q'$ is reached in which
$b$ is still globally ready but not anymore of maximal priority. Such a situation can be statically detected
and eliminated by additional priorities. We consider only specifications without this kind of confusion.

%% file: description.tex
\vspace{-0.5\baselineskip}
\subsection{Problem description}
\label{subsec: problem description}
A system $(S,<)$ is defined by a {\em composed system} $S$ --- of set of processes $P_i$, a set of
interactions and a priority order $<$ to be enforced, our goal is to
define a distributed implementation for $(S,<)$.  We
define an algorithm which
constructs such a distributed implementation by defining for each process a
{\em local controller} such that the joint execution of all processes $P_i$ and
their corresponding controllers guarantee the following:
\begin{enumerate} 
\item Any sequentialisation of an execution of the obtained concurrent execution is
      an execution of $(S,<)$, that is executions of $S$ respecting $<$
\item if $(S,<)$ is deadlock free, then no execution will deadlock
\end{enumerate}
Sequentialisations are obtained by arbitrarily ordering {\em concurrently} executed interactions.
Controllers are described as protocols interacting amongst each other by messages. Our aim is for each
process $P_i$ to be able to execute a next transition as quickly as possible, and not to minimise the 
number of messages sent.

%% file: BIPImplementation.tex
In this section, we provide here a description of the protocol used by each {\em local} controller 
associated to each process in a system $(S,<)$, which can be found in \cite{TRJLAP}. 
Each system has a fixed number of processes $P_i$. 
We consider here only binary interactions.

The protocol performs communication by messages exchanged 
between processes so as to be able to decide about a  next interaction to be fired jointly.
We also illustrate how our protocol behaves on a simple example.
We also assume that the internal activities of processes are terminating.
As quite usually, we assume that the message passing mechanism ensures the following basic properties:
1) any message is received at the destination within a finite delay;
2) messages sent from location $L_1$ to $L_2$ are received in the order in which they have been sent;
3) there is no duplication nor spontaneous creation of messages.
\subsection{Description of the protocol}
\label{subsec: Protocol}

We now describe the controllers of individual processes
which enforce correct executions, that is joint executions of synchronizations 
 and adherence to the global priority order. 
It is understood that what we call in the following ``process'' is in fact a {\em controlled process}
obtained by executing the process and its controller in the context of all peer processes.

For each interaction $a$ involved in at least one priority rule, one of the involved processes
$P_i$ plays the role of the negotiator for $a$.
If there exists at least one  interaction with higher
priority, the role of the negotiator is to check for the enablednes of $a$, and
if there exists at least one  interaction with lower
priority, its role is to answer readiness requests.

The Controller associated with each process, maintains a set of data structures shared and maintained
by the different subtasks of the controller:  $readySet$ (resp. $enabledSet$) contains the set
of  interactions which are known to be globally ready (resp. enabled) in the current local state $q$
 and $possibleSet$ maintains the set of interactions that are locally ready. Note that $possibleSet$
contains purely local information which can be calculated immediately when entering a new local state.
The other two sets are calculated by a series of message exchanges, and the complete information is 
generally not calculated but as soon as an interaction is known to be enabled, its triggering will be initiated.

\begin{figure}[H]
\begin{center}
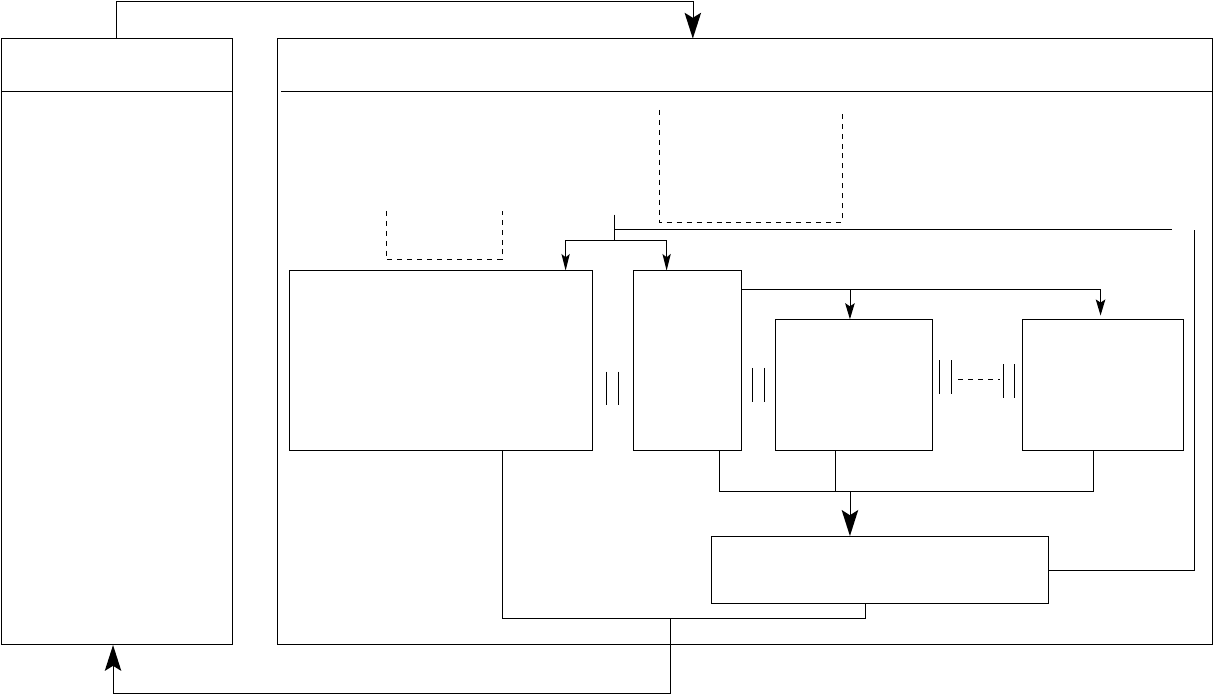    
\centering
\caption{Structure of the protocol for one process}      
\label{fig:protocol3}  
\end{center}                                         
\end{figure}

The general structure of the controller for each individual process $P_i$ is shown in
Figure~\ref{fig:protocol3}. The overall controller --- and the process to be controlled --- 
are represented as a set of parallel activities which we call threads, and which in our implementation are 
realized by Java threads (see Section~\ref{sec: expriments}) with a shared memory space and a set of shared 
message buffers. The detailed algorithms of these threads are given in an appendix at the end of the paper.
\vspace{-0.5\baselineskip}
\vspace{-0.5\baselineskip}
Indeed, incoming messages are stored until one of the activities
is ready to handle them. We use several FIFO buffers which are chosen such that the order amongst
messages stored in different buffers does not influence the algorithm; in particular, they are used
by concurrent threads.
 A buffer, which is read only by the 
thread $Main$, stores messages of the form
$\POSSIBLE(a)$, $\NOTPOSSIBLE(a)$, $\READY(a)$, $\NOTREADY(a)$,
and $\REFUSE(a)$. A second buffer stores 
messages of the form $\COMMIT(a)$, this buffer is read first by thread
$WaitingForCommit$, then by $TryToCommit$. 

The role of each message is
described in Table~\ref{Msgtable}. Given that we are handling binary interactions, 
 we do not explicitly mention the recipient nor the sender.
\vspace{-0.5\baselineskip}
\begin{table}[h]
\centering 
\begin{tabular}{c c c c} %
\hline 
&Message && Description\\ 
\hline\\
&\multicolumn{0}{l}{$\POSSIBLE$} && \multicolumn{0}{l}{Offer an interaction (which is locally ready)}\\ 
&\multicolumn{0}{l}{$\NOTPOSSIBLE$} && \multicolumn{0}{l}{respond that 
an interaction is not locally ready}\\ 

&\multicolumn{0}{l}{$\READY$} && \multicolumn{0}{l}{Ask about the global readiness of an interaction}\\
&\multicolumn{0}{l}{$\NOTREADY$} &&   \multicolumn{0}{l}{Respond that 
an interaction is not globally ready}\\
&\multicolumn{0}{l}{$\COMMIT$}  &&   \multicolumn{0}{l}{Commit to an interaction (cannot be undone by $P_i$)} \\

&\multicolumn{0}{l}{$\REFUSE$}  && \multicolumn{0}{l}{Inform that a process cannot commit to an interaction}\\
\hline \\ 
\end{tabular}

\caption{Messages used by the algorithm}
\label{Msgtable}
\end{table} 
\vspace{-0.5\baselineskip}

The controller $C_i$ associated with $P_i$ is either in state $Ready$ or in state $Busy$. 
In state $Busy$, $P_i$ executes the local action corresponding to the interaction chosen
by the protocol. Incoming messages are stored and will not be handled until
 the controller moves
into state $Ready$. 
In state $Ready$, $C_i$ looks for a next interaction to fire,
proceeding as follows: \smallskip

\begin{itemize}
 \item The $Main$ thread starts by checking its locally ready
   interactions ($possibleSet$) for interactions that are globally ready (see Algorithm~\ref{Main}). 
  To check the global readiness of an interaction $a$, messages of the form  $\POSSIBLE(a)$ are exchanged, and
  peers in which $a$ is currently not locally enabled respond with $\NOTPOSSIBLE(a)$ after which the requesting
  controller ``abandons'' $a$ until the process changes its state or the peer enters a state in which 
  $a$ is locally enabled  and sends a $POSSIBLE(a)$.

  Whenever it is detected that an interaction $a$ for which it plays the role of a negotiator is globally ready,
  a thread $Negotiate(a)$ is created which checks whether $a$ is enabled.
  If an interaction with maximal priority is globally ready, it is immediately known to be enabled
  and a COMMIT phase is entered (see further down).

\item the $Negotiate(a)$ thread checks the enabledness of an interaction $a$ (see Algorithm~\ref{negotiate}). 
 By sending a $READY(b)$ message to all negotiators of interactions
 $b$ with higher priority than $a$, it checks whether their interactions are globally ready 
  (and thus $a$ cannot be executed now).

 In turn the negotiators of $b$, as soon as they are not $BUSY$ and have found out whether $b$ is globally
 ready, respond positively or negatively as soon as they have the information available\footnote{In 
 fact, it is sufficient that $NONREADY(b)$ messages are sent as $a$ is blocked anyway as long as it does not
 have a response concerning $b$.}.

\item The $Main$ thread handles local priorities locally. Whenever an interaction
$b$ is known to be globally ready, it kills all threads $Negotiate(a)$ if $a<b$.

\item Concurrently to  $Main$, the thread $WaitingForCommit$ handles
  incoming $\COMMIT$ messages (see Algorithm~\ref{waitingForCommit}). Whenever
  a $\COMMIT(a)$ is received --- which means
  that $a$ is enabled and that the local process should commit to it\footnote{the existence 
  of the thread $WaitingForCommit$ means that no other actions is in its commit phase yet} --- 
  all other negotiation
  activities are terminated and a response $\COMMIT(a)$ is sent back to the peer.

\item As long as no commit phase is initiated by a peer, $Main$ tries to commit to the first interaction 
 found enabled
 (as a way to handle local conflicts) by activating $TryToCommit$. 
    $WaitingForCommit$ terminates once $TryToCommit$
is activated, to avoid multiple commits in the same state of $P_i$.

\item $TryToCommit(a)$ sends a $\COMMIT(a)$
  message to the corresponding peer and waits for a response (see Algorithm~\ref{tryToCommit}).
Note that if $TryToCommit$ fails committing to $a$ because it receives a $REFUSE$ message --- in that case
  the peer has committed to a conflicting interaction --- the process 
starts again by checking the global readiness of its locally ready interactions. Indeed, 
as the peer has committed to another action its state may have changed.
For the interactions $a$ for which there exists at least one interaction with higher priority,
the commit procedure is always initiated by the negotiator of $a$ who is the first one to know 
about $a$'s enabledness.

\item Finally, the thread $AnswerNegotiators$ is always active if the process $P_i$ is the negotiator 
  for at least one interaction $a$ that dominates some other interaction. 
 This thread receives messages of the form $\READY(a)$.
  It  returns 
  $\NOTREADY(a)$ if $a$ is in the $notReadySet$,
  and otherwise defers the answer until the status of $a$ is known. 
\end{itemize}

%% file: protocolOverviewF.pdf_t
\begin{picture}(0,0)%
\includegraphics{protocolOverviewF.pdf}%
\end{picture}%
\setlength{\unitlength}{1575sp}%
\begingroup\makeatletter\ifx\SetFigFontNFSS\undefined%
\gdef\SetFigFontNFSS#1#2#3#4#5{%
  \reset@font\fontsize{#1}{#2pt}%
  \fontfamily{#3}\fontseries{#4}\fontshape{#5}%
  \selectfont}%
\fi\endgroup%
\begin{picture}(14604,8349)(-1271,-6913)
\put(11116,-3211){\makebox(0,0)[lb]{\smash{{\SetFigFontNFSS{5}{6.0}{\rmdefault}{\mddefault}{\updefault}{\color[rgb]{0,0,0}\normalsize{Negotiate}}%
}}}}
\put(8191,-5506){\makebox(0,0)[lb]{\smash{{\SetFigFontNFSS{6}{7.2}{\rmdefault}{\mddefault}{\updefault}{\color[rgb]{0,0,0}\normalsize{TryToCommit}}%
}}}}
\put(6751,-916){\makebox(0,0)[lb]{\smash{{\SetFigFontNFSS{5}{6.0}{\rmdefault}{\mddefault}{\updefault}{\color[rgb]{0,0,0}NOTPOSSIBLE}%
}}}}
\put(6751,-601){\makebox(0,0)[lb]{\smash{{\SetFigFontNFSS{5}{6.0}{\rmdefault}{\mddefault}{\updefault}{\color[rgb]{0,0,0}REFUSE}%
}}}}
\put(6751,-1186){\makebox(0,0)[lb]{\smash{{\SetFigFontNFSS{5}{6.0}{\rmdefault}{\mddefault}{\updefault}{\color[rgb]{0,0,0}POSSIBLE}%
}}}}
\put(6751,-331){\makebox(0,0)[lb]{\smash{{\SetFigFontNFSS{5}{6.0}{\rmdefault}{\mddefault}{\updefault}{\color[rgb]{0,0,0}READY}%
}}}}
\put(6751,-61){\makebox(0,0)[lb]{\smash{{\SetFigFontNFSS{5}{6.0}{\rmdefault}{\mddefault}{\updefault}{\color[rgb]{0,0,0}NOTREADY}%
}}}}
\put(3556,-1546){\makebox(0,0)[lb]{\smash{{\SetFigFontNFSS{5}{6.0}{\rmdefault}{\mddefault}{\updefault}{\color[rgb]{0,0,0}COMMIT}%
}}}}
\put(6526,-2851){\makebox(0,0)[lb]{\smash{{\SetFigFontNFSS{5}{6.0}{\rmdefault}{\mddefault}{\updefault}{\color[rgb]{0,0,0}\normalsize{Main}}%
}}}}
\put(8146,-3211){\makebox(0,0)[lb]{\smash{{\SetFigFontNFSS{5}{6.0}{\rmdefault}{\mddefault}{\updefault}{\color[rgb]{0,0,0}\normalsize{Negotiate}}%
}}}}
\put(6076,524){\makebox(0,0)[lb]{\smash{{\SetFigFontNFSS{8}{9.6}{\rmdefault}{\mddefault}{\updefault}{\color[rgb]{0,0,0}\normalsize{READY}}%
}}}}
\put(-674,524){\makebox(0,0)[lb]{\smash{{\SetFigFontNFSS{8}{9.6}{\rmdefault}{\mddefault}{\updefault}{\color[rgb]{0,0,0}\normalsize{BUSY}}%
}}}}
\put(-1124,-2806){\makebox(0,0)[lb]{\smash{{\SetFigFontNFSS{5}{6.0}{\rmdefault}{\mddefault}{\updefault}{\color[rgb]{0,0,0}\normalsize{Execution of}}%
}}}}
\put(-1124,-3256){\makebox(0,0)[lb]{\smash{{\SetFigFontNFSS{5}{6.0}{\rmdefault}{\mddefault}{\updefault}{\color[rgb]{0,0,0}\normalsize{the interaction}}%
}}}}
\put(2296,-2851){\makebox(0,0)[lb]{\smash{{\SetFigFontNFSS{5}{6.0}{\rmdefault}{\mddefault}{\updefault}{\color[rgb]{0,0,0}\normalsize{WaitingForCommit}}%
}}}}
\end{picture}%

%% file: cycles.tex
\vspace{-\baselineskip}
\subsection{Decision cycles}
\label{cycles}
In order to avoid deadlocks due to decision cycles amongst interactions in conflict,
we introduce a notion of {\em cycle} representing potential decision cycles.
We denote by {\em inter($a$,P$_1$,P$_2$)} the fact that interaction $a$  involves  processes P$_1$ and P$_2$.
A {\em cycle}, $C_A$ is a set of interactions $A=\{a_i\}^n_{i=1}$ for which the following holds:
there exist $n$ processes $\{P_i\}^n_{i=1}$, such that 
 {$\bigwedge_{i=1}^{n}$\em inter($a_i$,P$_i$,P$_{i+1\, mod\, n}$)}.
In addition, we require that there exists at least one global state in which 
all conflicting interactions are enabled.
A {\em cycle} $C_A$ bears indeed a risk of deadlock or livelock in 
a state in which all interactions of $C_A$ are enabled. Indeed, it 
represents a symmetric situation for all involved processes, where each process could
wait forever a response to a $COMMIT$ (deadlock) or propose a choice of a next interaction
representing a differnet solution than the one chosen (locally) by all others, reject it by sending 
a $REFUSE$ and then start all over again forever (livelock). This is a well-known problem in the 
context of communicating processes.
In~\cite{Bagrodia89} a total order over the system interactions is defined, which allows to avoid deadlock by 
executing the interaction with higher order if an actual conflict occurs. In~\cite{PerezCT04}, 
a similar solution is proposed by  imposing a total order over all processes, which breaks the cycle 
by always executing the interaction proposed by the process with higher order.

The solution we propose is to detect statically the set of (minimal) cycles of 
the system. Then, in a second step, we define for  each cycle statically a process of the cycle playing the
role of the {\em Cyclebreaker}.
This particular process will arbitrate when a blocking situation actually occurs. 
This approach avoids defining a total order of all interactions or processes which is 
useless if there is no cycle. \vspace{-0.5\baselineskip}

\paragraph{Illustrative example}

\begin{center}
  \begin{figure}
     \begin{center}
       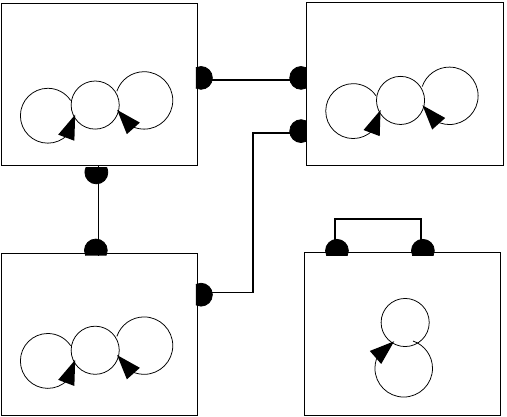
       \hspace{0.7cm}
       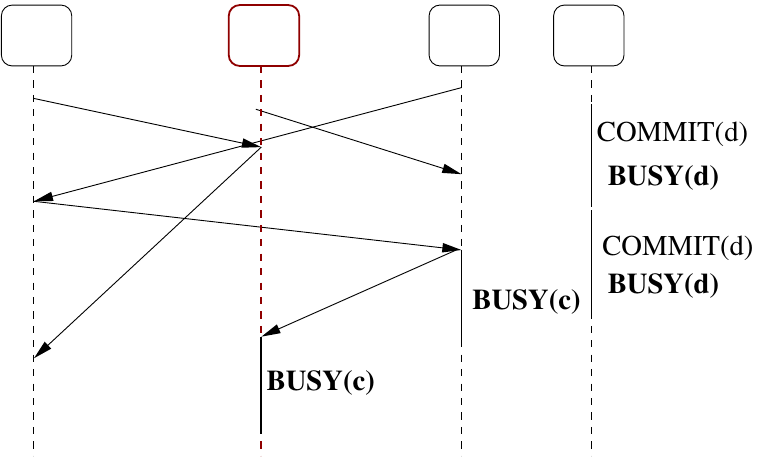
        \end{center}    
    \caption{An example with cycle and independence}\label{fig:cycleExample}
  \end{figure}
\end{center}

 Figure \ref{fig:cycleExample} depicts an example representing a cycle.
The system consists of 4 components: 3 processes
 $\{P_1, P_2, P_3\}$ forming a cycle $C_A$ for the set of interactions $A=\{a, b, c\}$,
 and a completely independent process $P_4$.
 The existence of a cycle can 
be concluded from the structure and the behaviors of the processes 
(the interactions $a$, $b$, $c$ are always enabled and in conflict).  If no priority rules
 are defined on the set of interactions $A$, then the algorithm --- as explained so far -- may
end in a deadlock.
A possible deadlock scenario is depicted on the right side of Figure~\ref{fig:cycleExample}. 
This occurs when $P_i$ sends a $COMMIT$ message to $P_{i+1}$ and waits
for it. Which means that each $P_i$ is waiting for a response from its peer who has made another choice
and is waiting as well.
According to the proposed solution, let us suppose that $P_2$ is chosen as the {\em Cyclebreaker} of
$C_A$. According to Algorithm~\ref{tryToCommit} (as described in Figure~\ref{fig:cycleExample}), whenever
process $P_i$ which is already engaged in committing an interaction and which receives a $COMMIT$ for
a different interaction, will send back a $REFUSE$ message only if the $COMMIT$ comes from a process 
which is not the {\em Cyclebreaker}. This breaks the cycle.
Independently, the process $P_4$ can perform whenever it is possible the interaction $d$.
We have proven the correctness of this algorithm in~\cite{TRJLAP}. In this paper, we present an implementation
and its use in a number of experiments.

%% file: cycleExample1.pdf_t
\begin{picture}(0,0)%
\includegraphics{cycleExample1.pdf}%
\end{picture}%
\setlength{\unitlength}{1948sp}%
\begingroup\makeatletter\ifx\SetFigFontNFSS\undefined%
\gdef\SetFigFontNFSS#1#2#3#4#5{%
  \reset@font\fontsize{#1}{#2pt}%
  \fontfamily{#3}\fontseries{#4}\fontshape{#5}%
  \selectfont}%
\fi\endgroup%
\begin{picture}(4906,4041)(2554,-6154)
\put(3781,-2716){\makebox(0,0)[lb]{\smash{{\SetFigFontNFSS{8}{9.6}{\rmdefault}{\bfdefault}{\updefault}{\color[rgb]{0,0,0}$a$}%
}}}}
\put(3286,-3211){\makebox(0,0)[lb]{\smash{{\SetFigFontNFSS{8}{9.6}{\rmdefault}{\mddefault}{\updefault}{\color[rgb]{0,0,0}$S_1$}%
}}}}
\put(2746,-2896){\makebox(0,0)[lb]{\smash{{\SetFigFontNFSS{8}{9.6}{\rmdefault}{\bfdefault}{\updefault}{\color[rgb]{0,0,0}$b$}%
}}}}
\put(3781,-5101){\makebox(0,0)[lb]{\smash{{\SetFigFontNFSS{8}{9.6}{\rmdefault}{\bfdefault}{\updefault}{\color[rgb]{0,0,0}$c$}%
}}}}
\put(3286,-5596){\makebox(0,0)[lb]{\smash{{\SetFigFontNFSS{8}{9.6}{\rmdefault}{\mddefault}{\updefault}{\color[rgb]{0,0,0}$S_2$}%
}}}}
\put(2746,-5281){\makebox(0,0)[lb]{\smash{{\SetFigFontNFSS{8}{9.6}{\rmdefault}{\bfdefault}{\updefault}{\color[rgb]{0,0,0}$b$}%
}}}}
\put(6751,-2671){\makebox(0,0)[lb]{\smash{{\SetFigFontNFSS{8}{9.6}{\rmdefault}{\bfdefault}{\updefault}{\color[rgb]{0,0,0}$a$}%
}}}}
\put(6256,-3166){\makebox(0,0)[lb]{\smash{{\SetFigFontNFSS{8}{9.6}{\rmdefault}{\mddefault}{\updefault}{\color[rgb]{0,0,0}$S_3$}%
}}}}
\put(5716,-2851){\makebox(0,0)[lb]{\smash{{\SetFigFontNFSS{8}{9.6}{\rmdefault}{\bfdefault}{\updefault}{\color[rgb]{0,0,0}$c$}%
}}}}
\put(6301,-5326){\makebox(0,0)[lb]{\smash{{\SetFigFontNFSS{8}{9.6}{\rmdefault}{\mddefault}{\updefault}{\color[rgb]{0,0,0}$S_4$}%
}}}}
\put(6886,-5821){\makebox(0,0)[lb]{\smash{{\SetFigFontNFSS{8}{9.6}{\rmdefault}{\bfdefault}{\updefault}{\color[rgb]{0,0,0}$d$}%
}}}}
\put(6050,-4150){\makebox(0,0)[lb]{\smash{{\SetFigFontNFSS{9}{10.8}{\rmdefault}{\bfdefault}{\updefault}{\color[rgb]{0,0,0}$d$}%
}}}}
\put(4817,-2797){\makebox(0,0)[lb]{\smash{{\SetFigFontNFSS{9}{10.8}{\rmdefault}{\bfdefault}{\updefault}{\color[rgb]{0,0,0}$a$}%
}}}}
\put(4636,-4156){\makebox(0,0)[lb]{\smash{{\SetFigFontNFSS{9}{10.8}{\rmdefault}{\bfdefault}{\updefault}{\color[rgb]{0,0,0}$c$}%
}}}}
\put(2656,-4921){\makebox(0,0)[lb]{\smash{{\SetFigFontNFSS{10}{12.0}{\rmdefault}{\bfdefault}{\updefault}{\color[rgb]{0,0,0}$P_2$}%
}}}}
\put(2656,-2491){\makebox(0,0)[lb]{\smash{{\SetFigFontNFSS{10}{12.0}{\rmdefault}{\bfdefault}{\updefault}{\color[rgb]{0,0,0}$P_1$}%
}}}}
\put(5581,-4966){\makebox(0,0)[lb]{\smash{{\SetFigFontNFSS{10}{12.0}{\rmdefault}{\bfdefault}{\updefault}{\color[rgb]{0,0,0}$P_4$}%
}}}}
\put(5626,-2491){\makebox(0,0)[lb]{\smash{{\SetFigFontNFSS{10}{12.0}{\rmdefault}{\bfdefault}{\updefault}{\color[rgb]{0,0,0}$P_3$}%
}}}}
\put(3061,-4201){\makebox(0,0)[lb]{\smash{{\SetFigFontNFSS{9}{10.8}{\rmdefault}{\bfdefault}{\updefault}{\color[rgb]{0,0,0}$b$}%
}}}}
\end{picture}%

%% file: cycleScenario.pdf_t
\begin{picture}(0,0)%
\includegraphics{cycleScenario.pdf}%
\end{picture}%
\setlength{\unitlength}{2279sp}%
\begingroup\makeatletter\ifx\SetFigFontNFSS\undefined%
\gdef\SetFigFontNFSS#1#2#3#4#5{%
  \reset@font\fontsize{#1}{#2pt}%
  \fontfamily{#3}\fontseries{#4}\fontshape{#5}%
  \selectfont}%
\fi\endgroup%
\begin{picture}(6345,3790)(2419,-4088)
\put(2791,-1861){\rotatebox{15.0}{\makebox(0,0)[lb]{\smash{{\SetFigFontNFSS{8}{9.6}{\rmdefault}{\mddefault}{\updefault}{\color[rgb]{0,0,0}COMMIT(a)}%
}}}}}
\put(6076,-601){\makebox(0,0)[lb]{\smash{{\SetFigFontNFSS{8}{9.6}{\rmdefault}{\mddefault}{\updefault}{\color[rgb]{0,0,0}$P_{3}$}%
}}}}
\put(2926,-2986){\rotatebox{43.0}{\makebox(0,0)[lb]{\smash{{\SetFigFontNFSS{8}{9.6}{\rmdefault}{\mddefault}{\updefault}{\color[rgb]{0,0,0}REFUSE(b)}%
}}}}}
\put(4726,-2131){\rotatebox{355.0}{\makebox(0,0)[lb]{\smash{{\SetFigFontNFSS{8}{9.6}{\rmdefault}{\mddefault}{\updefault}{\color[rgb]{0,0,0}REFUSE(a)}%
}}}}}
\put(4816,-1186){\rotatebox{343.0}{\makebox(0,0)[lb]{\smash{{\SetFigFontNFSS{8}{9.6}{\rmdefault}{\mddefault}{\updefault}{\color[rgb]{0,0,0}COMMIT(c)}%
}}}}}
\put(4681,-2986){\rotatebox{24.0}{\makebox(0,0)[lb]{\smash{{\SetFigFontNFSS{8}{9.6}{\rmdefault}{\mddefault}{\updefault}{\color[rgb]{0,0,0}COMMIT(c)}%
}}}}}
\put(4411,-601){\makebox(0,0)[lb]{\smash{{\SetFigFontNFSS{8}{9.6}{\rmdefault}{\bfdefault}{\updefault}{\color[rgb]{.56,0,0}$P_{2}$}%
}}}}
\put(7111,-601){\makebox(0,0)[lb]{\smash{{\SetFigFontNFSS{8}{9.6}{\rmdefault}{\mddefault}{\updefault}{\color[rgb]{0,0,0}$P_{4}$}%
}}}}
\put(2971,-1051){\rotatebox{348.0}{\makebox(0,0)[lb]{\smash{{\SetFigFontNFSS{8}{9.6}{\rmdefault}{\mddefault}{\updefault}{\color[rgb]{0,0,0}COMMIT(b)}%
}}}}}
\put(2521,-601){\makebox(0,0)[lb]{\smash{{\SetFigFontNFSS{8}{9.6}{\rmdefault}{\mddefault}{\updefault}{\color[rgb]{0,0,0}$P_{1}$}%
}}}}
\end{picture}%

%% file: implementation.tex
\begin{figure}[H]
\begin{center}
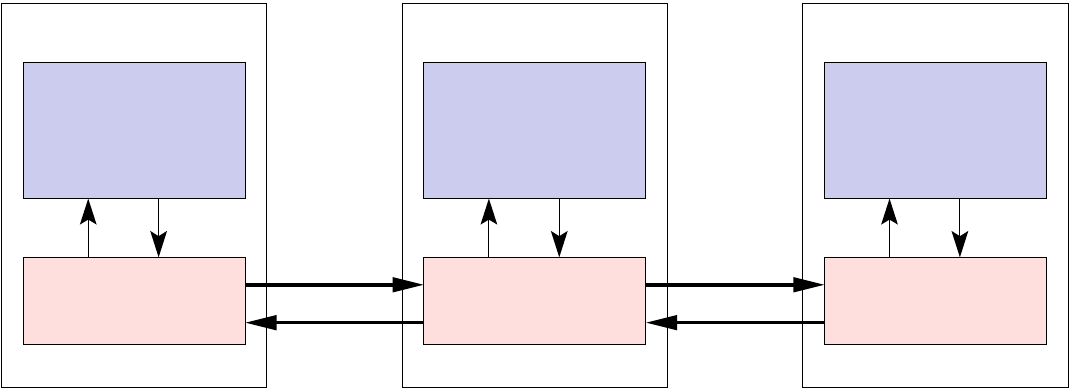    
\centering
\caption{Implementation layers}      
\label{fig:implem}  
\end{center}                                    
\end{figure}

We have implemented the protocol described in Section~\ref{sec: protocol}
 using Java~1.6 and Message Passing Interfaces (MPI) in order to
experiment its efficiency on examples of different nature.
We have used the MPI 
 library \cite{SnirOttoHussLedermanEtAl96} to perform the communication layer of our algorithm 
because of its good performance, usage facility and its portability \cite{MPIPortable}. 
In our prototype, the exchange of messages between processes
is performed at the MPI layer and all the computation operations of our algorithm 
are performed at the Java program level (see Figure~\ref{fig:implem}).
In this section, we show how we have evaluated the performance of our algorithm on hand of  the implemented prototype.
Tests have been run on a set of $2.2$ GHz Intel machines with $2$ GB RAM, in a configuration
where  each physical machine hosts only one process. We have however not made sure that no other application 
is running during the experiments, just that the overall charge on each machine is ``low''.

Our experiments evaluated essentially two metrics which are comparable to those used also in \cite{PerezCT04}: 
the first is a metric called {\em message-count} 
which measures the (average) number of messages required to schedule an interaction for execution, starting
from the moment on that it is ready in one of the involved processes. 
The second one is called {\em Response-Time} and is defined as the sum of two other metrics
 {\em Sync-Time} and {\em selection-Time}:
 {\em Sync-Time} measures the (mean) time taken by the algorithm to ensure that a given interaction
is globally {\em ready}, again starting from the moment where it is locally ready in at least one of the 
peers\footnote{an alternative option would be to measure only from the moment on where the
interaction is already enabled, that is only the time required to "detect" this enabledness; this is however
quite difficult to evaluate in a distributed setting.}. 
{\em Selection-Time} measures the (mean) time taken by the algorithm to select an interaction 
for execution once it has been found enabled.

All metrics are measured for a given system by experimenting with different choices of 
parameters. We then analyze how variations of parameters affect the considered metrics
and compare them to theoretical analysis on the algorithm. 

We also compare for an example without priorities the
 {\em message-count} metric obtained for our algorithm and for an implementation
 of the $\alpha$-core algorithm. We could not compare execution times because the implementation
of $\alpha$-core we have at hand cannot be run in the same setting and the data provided in
\cite{PerezCT04} are obtained in a incomparable setting as well.

\vspace{-0.5\baselineskip}
\vspace{-0.5\baselineskip}

%% file: implemOverview.pdf_t
\begin{picture}(0,0)%
\includegraphics{implemOverview.pdf}%
\end{picture}%
\setlength{\unitlength}{2279sp}%
\begingroup\makeatletter\ifx\SetFigFontNFSS\undefined%
\gdef\SetFigFontNFSS#1#2#3#4#5{%
  \reset@font\fontsize{#1}{#2pt}%
  \fontfamily{#3}\fontseries{#4}\fontshape{#5}%
  \selectfont}%
\fi\endgroup%
\begin{picture}(8889,3219)(2014,-3988)
\put(5626,-1906){\makebox(0,0)[lb]{\smash{{\SetFigFontNFSS{10}{12.0}{\rmdefault}{\mddefault}{\updefault}{\color[rgb]{0,0,0}Java-Instance$_1$}%
}}}}
\put(5806,-3346){\makebox(0,0)[lb]{\smash{{\SetFigFontNFSS{7}{8.4}{\rmdefault}{\mddefault}{\updefault}{\color[rgb]{0,0,0}\normalsize{MPI-rank$_1$}}%
}}}}
\put(8956,-1906){\makebox(0,0)[lb]{\smash{{\SetFigFontNFSS{10}{12.0}{\rmdefault}{\mddefault}{\updefault}{\color[rgb]{0,0,0}Java-Instance$_2$}%
}}}}
\put(9136,-3346){\makebox(0,0)[lb]{\smash{{\SetFigFontNFSS{7}{8.4}{\rmdefault}{\mddefault}{\updefault}{\color[rgb]{0,0,0}\normalsize{MPI-rank$_2$}}%
}}}}
\put(2296,-1906){\makebox(0,0)[lb]{\smash{{\SetFigFontNFSS{10}{12.0}{\rmdefault}{\mddefault}{\updefault}{\color[rgb]{0,0,0}Java-Instance$_0$}%
}}}}
\put(2476,-3346){\makebox(0,0)[lb]{\smash{{\SetFigFontNFSS{7}{8.4}{\rmdefault}{\mddefault}{\updefault}{\color[rgb]{0,0,0}\normalsize{MPI-rank$_0$}}%
}}}}
\put(2116,-1096){\makebox(0,0)[lb]{\smash{{\SetFigFontNFSS{9}{10.8}{\rmdefault}{\mddefault}{\updefault}{\color[rgb]{0,0,0}P$_0$}%
}}}}
\put(5491,-1096){\makebox(0,0)[lb]{\smash{{\SetFigFontNFSS{9}{10.8}{\rmdefault}{\mddefault}{\updefault}{\color[rgb]{0,0,0}P$_1$}%
}}}}
\put(8776,-1096){\makebox(0,0)[lb]{\smash{{\SetFigFontNFSS{9}{10.8}{\rmdefault}{\mddefault}{\updefault}{\color[rgb]{0,0,0}P$_2$}%
}}}}
\end{picture}%

%% file: degreeConflict.tex
\vspace{-1.5\baselineskip}
\subsection{Sensitivity to the degree of conflict}
\vspace{-0.5\baselineskip}
First, we study the sensitivity of our algorithm to the degree
 of conflict in a given system. 
The degree of conflict ($d$) is measured by the number 
of interactions that may be in actual conflict with any (or a particular) interaction.
Remember that we distinguish between  {\em structural} and
{\em prioritized} conflict (see Definition~\ref{prioconflict}). 
\vspace{-\baselineskip}
\subsubsection{Sensitivity to prioritized conflicts}
\vspace{-0.5\baselineskip}
The purpose of our algorithm is to ensure correct synchronization between processes 
by respecting global priorities. We first show some results concerning
 {\em prioritized} conflicts. 
\vspace{-0.5\baselineskip}
\begin{figure}[H]
\begin{center}
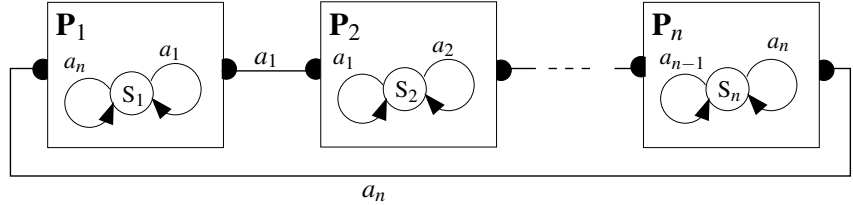    
\centering
\caption{System pattern for experiments}      
\label{fig:cycleN}  
\end{center}                                       
\end{figure}
\vspace{-0.5\baselineskip}
An evaluation has been undertaken using the example 
depicted in Figure~\ref{fig:cycleN} with a single global state. For each considered configration, the system has been executed
several times, and each execution has been terminated at the execution of the first interaction.
 The system consists of a set of $n$ processes, 
and a set of $n$ interactions building a circular chain.
 This pattern is flexible and it allows as to observe 
how our algorithm performs in different situations. In fact, we can easily add both 
local and global priorities.\smallskip

Considering a given system, that is a composition $\|(P_1,...,P_n)$, $d$ can be increased 
by adding priority constraints. Here, we simply count the maximal number of priorities in 
which a single process is involved, in order to obtain the 
degree $d$, but as the discussion will reveal, finer measures could also be considered.
interactions of different processes of $S$ than $<_2$.\smallskip

As already explained, our experiments are performed on a system as depicted in 
Figure~\ref{fig:cycleN}, for $n=4$ and using the following priorities to achieve 
different degrees of conflict, where process $P_2$ --- which is chosen as the negotiator of $a_1$ ---
 is the process which in all cases is involved in all the priorities, whereas other processes are involved in at most
two of them: $d=0$: no priorities, $d=1$: $a_2<a_1$, $d=2$: $a_2<a_1\ \wedge a_3<a_1$, 
$d=3$: $a_2<a_1\ \wedge a_3<a_1\ \wedge a_4<a_1$, $d=4$: $a_2<a_1\ \wedge a_3<a_1\ \wedge a_4<a_1\ \wedge a_3<a_2$.
We have measured the average {\em message-count}, {\em response-time}, {\em sync-time}
 and the {\em selection-time} for all for cases.
\begin{center}
  \begin{figure}
    \begin{center}
     \includegraphics[width=0.4\textwidth, height=0.36\textheight, angle=-90]{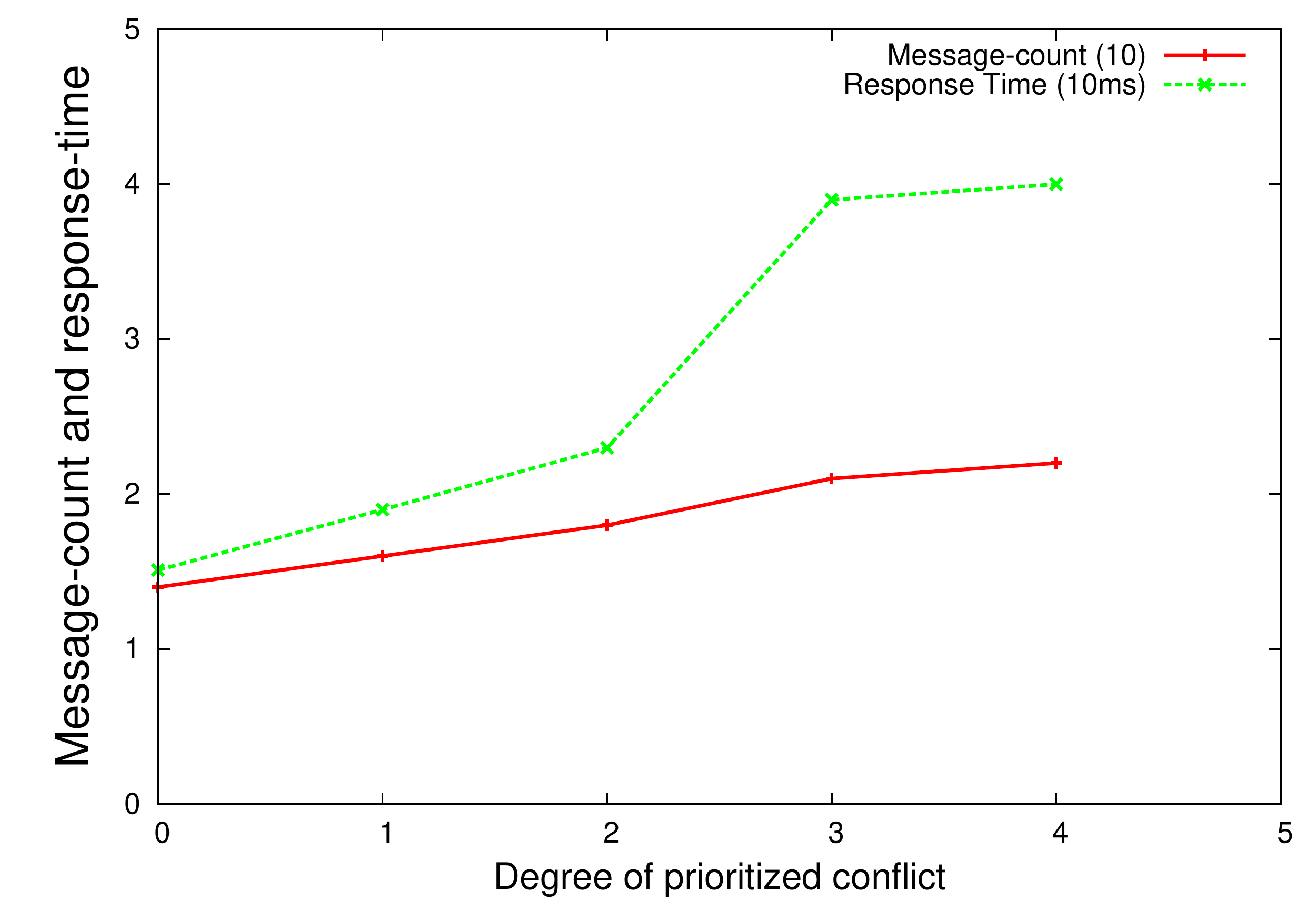}
      \hspace{0.03cm}
     \includegraphics[width=0.4\textwidth, height=0.36\textheight, angle=-90]{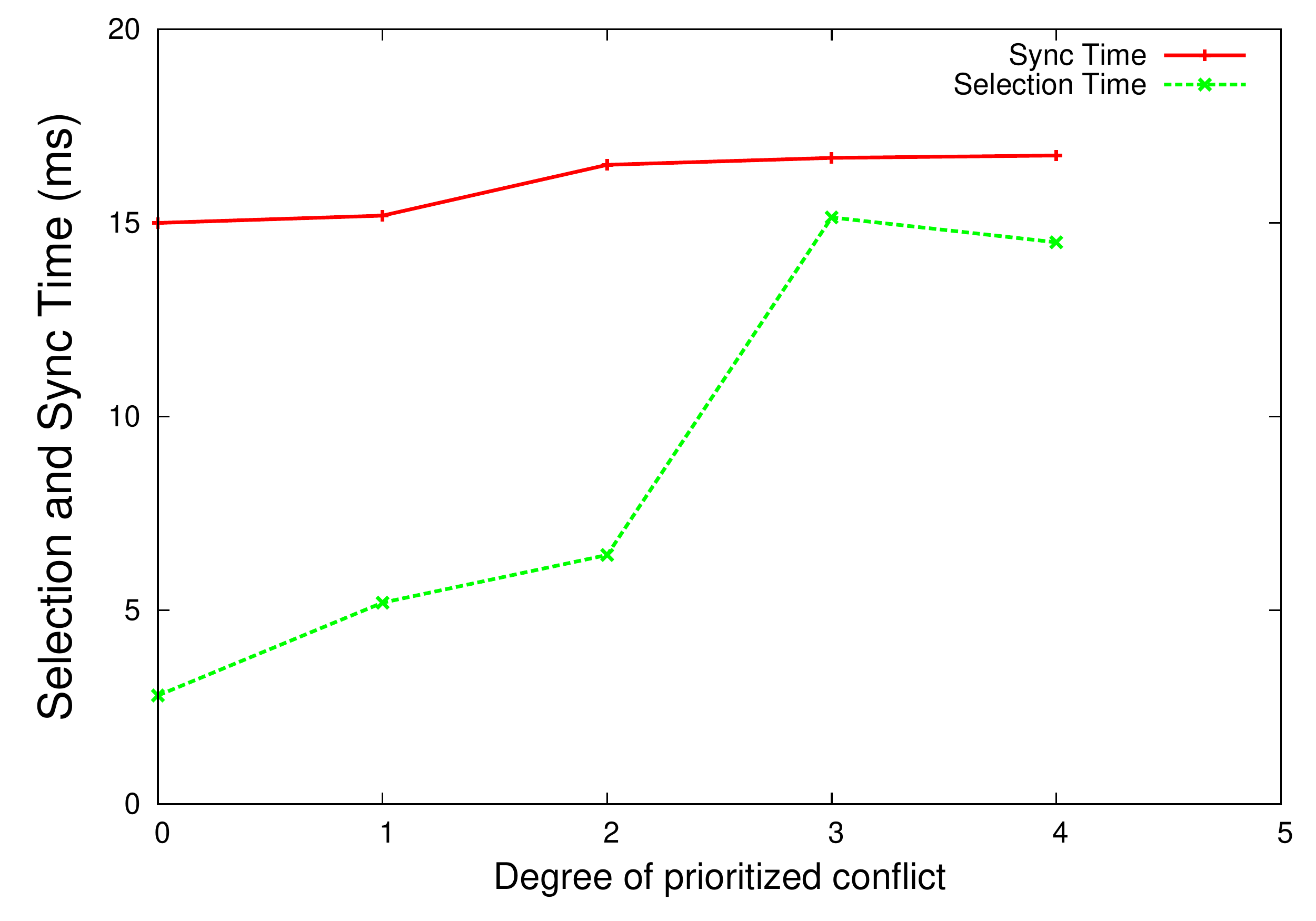}
     \end{center}    
    \caption{Sensitivity to the degree of {\em prioritized} conflict}\label{fig:philoPb}
\label{fig: priorityconflict}
  \end{figure}
\vspace{-0.5\baselineskip}
\end{center}
\paragraph{Variation of metric message-count}
Figure~\ref{fig: priorityconflict} shows that --- as expected --- the number of messages exchanged in order to execute 
the first (and unique) interaction
 increases with the degree $d$ of the system. 
Increasing $d$ means that more interactions are involved 
in priority rules, and thus more messages of type $READY$ are exchanged, and globally less interactions can be executed. 
\smallskip
In the case chosen for $d=1$, the  priority is defined by the rule $a_2<a_1$ which is a {\em local} priority involving
only process $P_2$. Thus, no negotiations and no $READY$ messages are needed which makes in principle, the same
 {\em message-count} as for $d=0$. This is confirmed by Figure~\ref{fig: priorityconflict} showing a non significant difference
between $d=0$ and $d=1$. For the case  $d=2$, the selected priorities
 are $a_2 < a_1$ and $a_3<a_1$. This means that the negotiator of $a_3$ ($P_3$) 
has to send a $READY$ message to  the negotiator of $a_1$ ($P_2$) and the latter has to send back as a response a $READY$ message which makes
 $2$ extra messages added comparing to the case of $d=0$. This is confirmed by the experimental results.\smallskip
\vspace{-\baselineskip}
\paragraph{Variation of metric response-time}
As expected, also the time required to execute the first interaction increases with the degree $d$ of the system, which can 
also be seen in Figure~\ref{fig: priorityconflict}. Again, adding a local conflict (as in the step from $d=0$ to $d=1$)
leads only to a small increase of the response time as the situation is handled locally. The increase is larger when a global 
priority is added. Note also that the increase in response time is more important than the increase of the number of messages:
up to 20
but indeed, adding a priority requires adding some explicit threads for negotiation, and on the system configuration
we use, the time is mainly spent for execution, whereas the communication time is relatively small. 
Figure~\ref{fig: priorityconflict} shows also the sensitivity of the {\em sync-time} and the {\em selection-time}  
of our prototype to the variation of  $d$.
Theoretically, the average synchronization time is independent of the number of conflicting interactions in our system.
Indeed, to decide the global readiness of a given interaction, a process has to send and receive 
a $POSSIBLE$ message for this interaction, which is completely independent of whether this interaction is involved or not 
in a priority rule. This is confirmed by the results of {\em sync-time} for $d=0$, $d=1$ and $d=2$ (given 
in Figure~\ref{fig: priorityconflict}), for which the synchronization time 
is almost the same.\smallskip
The synchronization times are slightly greater in case $d=3$ and $d=4$. 
This is due to the order in which messages are received. More precisely, for $d=2$
 priorities are $a_2<a_1$ and $a_3<a_1$ which implies that the process negotiating $a_3$ will 
send a $READY$ message to the negotiator of $a_1$ to check its readiness. Thus, the negotiator of 
$a_1$ may receive and treat this $READY$ message before reacting to the $POSSIBLE$ messages for the other interaction.
\smallskip
We can observe however that for increasing $d$, the time required to actually choose an enabled interaction, increases
considerably. This is not surprising. The fact that the selection time remains relatively small with respect to the synchronization
time allows  the overall response time increase to remain moderate.
\vspace{-0.5\baselineskip}
\subsubsection{Sensitivity to structural conflicts}
\begin{figure}[H]
\begin{center}
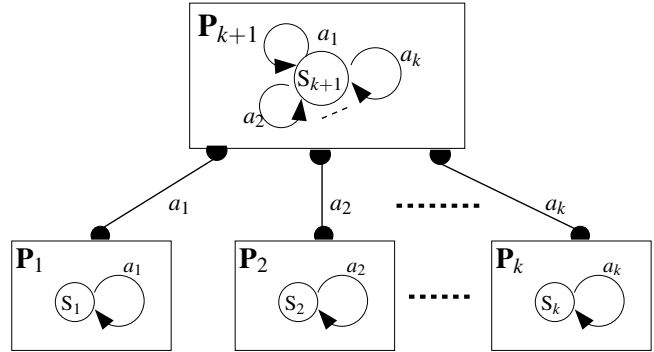    
\centering
\caption{System pattern for experiments ($T_k$)}      
\label{fig:Tsystem}  
\end{center}                                       
\end{figure}

\vspace{-0.5\baselineskip}
{\em Structural} conflict arises  between interactions when they are all in the $possibleSet$ of a common process.
To study how our algorithm performs with an increasing number of structural conflicts, we have carried out a series of 
experiments on a system as depicted in Figure~\ref{fig:Tsystem}.
We use a set of systems  $T_1,T_2, ..., T_n$ where each $T_k$ has $k$ binary interactions, 
referred to as $a_i$ ($i=1,2, ...,k$), and $k+1$ processes, referred to as $P_i$ ($i=1,2, ...,k+1$). Processes
$P_i$ participate in interaction $a_i$,  and $P_{k+1}$
participates in any interaction. Therefore, all interactions are in {\em structural} conflict, and the degree of the 
structural conflict can be measured by the number of processes in the system.
\vspace{-0.5\baselineskip}
\vspace{-0.5\baselineskip}
Each experiment consisted in executing 100 interactions, and we have evaluated our metrics for up to 5 conflicting interactions 
(a system with six processes) for several executions of this experiment for each degree of structural conflict.
\vspace{-\baselineskip}
\paragraph{Variation of message-count} We can see in the left side of  Figure~\ref{fig: strucutralconflict} that our algorithm requires 
considerably less messages than  $\alpha$-core, where we compare with the numbers provided in \cite{PerezCT04}
for this same  example. This is due to the fact that 
$\alpha$-core is ``connector-centric'', that is,
it creates an additional process for each interaction whereas our algorithm is process centric, that is
all negotiations are hosted by some process and share the same memory space. This means that our algorithm
can exploit more  local ``knowledge'' to execute interactions which reduces the number of messages exchanged.
When there is no conflict at all (in $T_1$) both algorithms exchange the same 
number of messages, then when the degree of conflict increases, our algorithm performs better.
The system $T_1$ has no conflict, and to execute $a_1$, $3$ messages are exchanged 
(one $POSSIBLE$ and two $COMMIT$), thus $300$ messages are transmitted during the experiment. 
When there are conflicts, for  $T_2$ for example, again $3$ messages
 are needed to execute an interaction in the best case, but every time an interaction is refused, 
at the worst case,  a penalty of $3$ messages is added (one $POSSIBLE$, one $COMMIT$ and one $REFUSE$). 
To execute $100$ interactions, $300$ messages are needed in the best case, and $212$ extra messages 
have been added for the situations where an interaction has been refused.
\begin{center}
 \vspace{-0.5\baselineskip}
  \begin{figure}
    \begin{center}
     \includegraphics[width=0.4\textwidth, height=0.36\textheight, angle=-90]{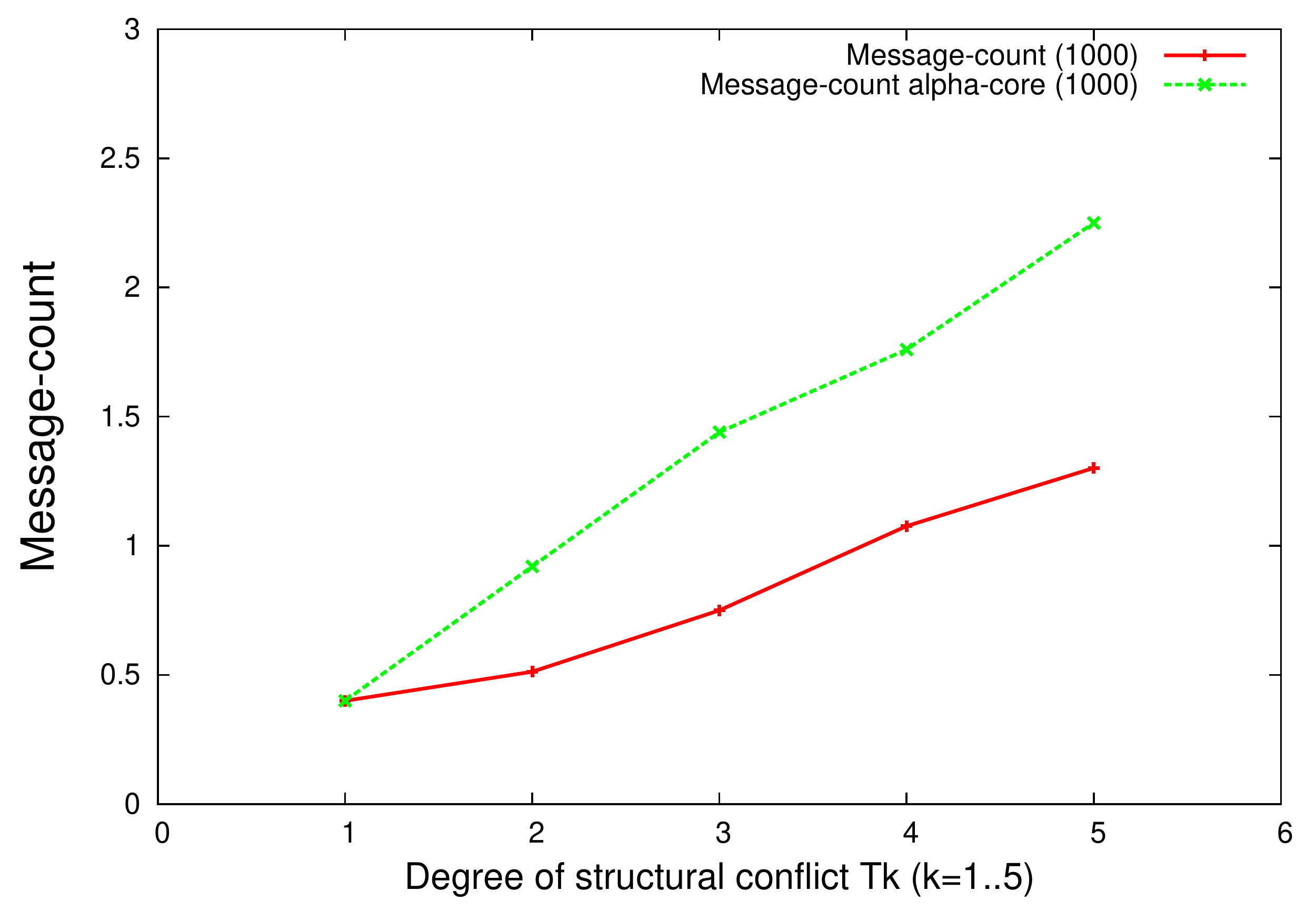}
      \hspace{0.03cm}
     \includegraphics[width=0.4\textwidth, height=0.36\textheight, angle=-90]{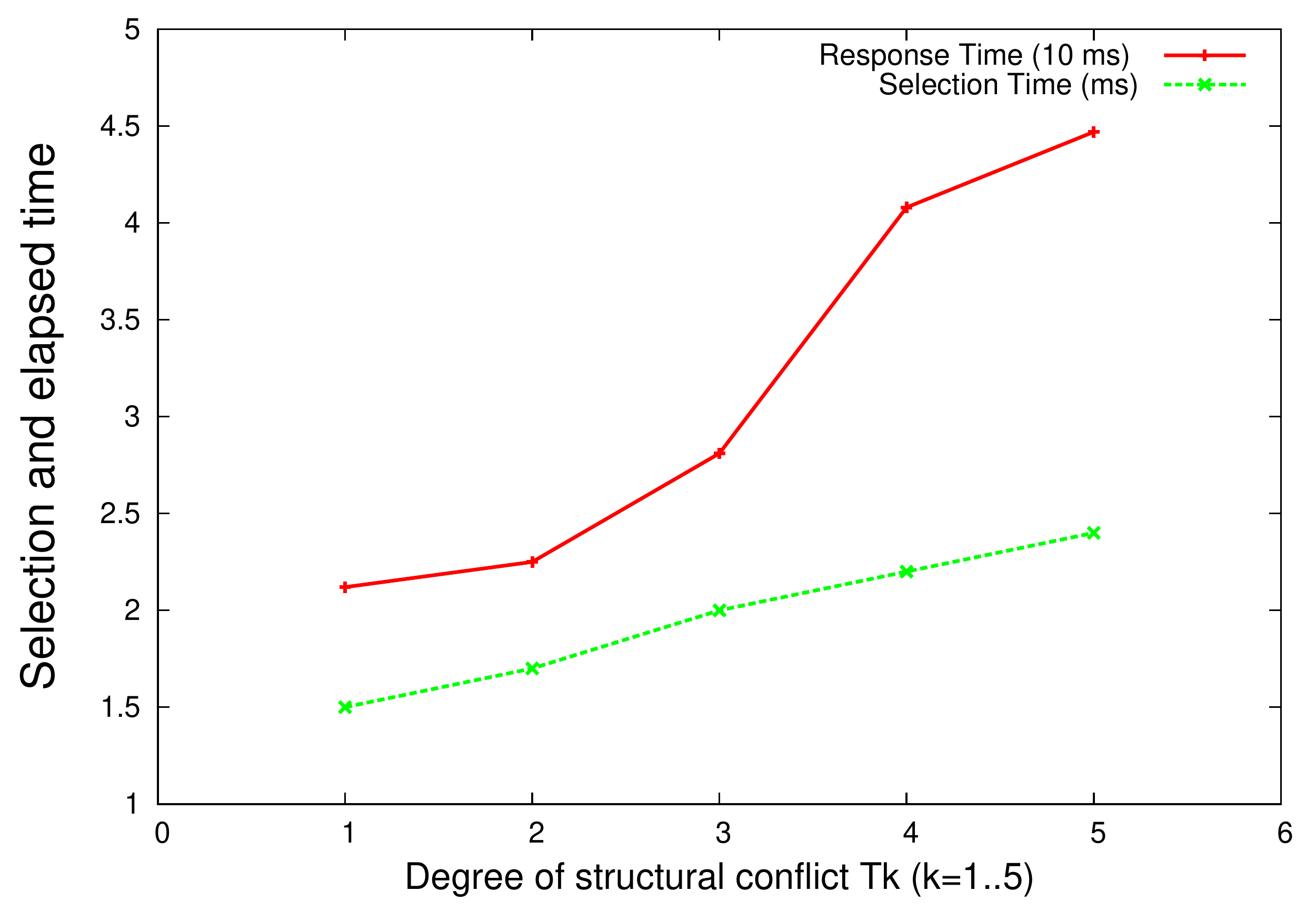}
   \end{center}    
     \caption{Sensitivity to the degree of {\em structural} conflict}
\label{fig: strucutralconflict}
  \end{figure}
\vspace{-0.5\baselineskip}
\end{center}
\vspace{-0.5\baselineskip}
\paragraph{Variation of response-time} Figure~\ref{fig: strucutralconflict} shows also the selection and elapsed time. 
Again, the average selection time is in principle independent of the number of interactions in {\em structural} conflict. 
Because, when no priorities are added and when an interaction becomes {\em ready}, only two $COMMIT$ messages 
are exchanged to execute an interaction. Thus the average selection time should be of about $2*\lambda$, where $\lambda$ is the average
 message transmission time which in our experimental architecture is  $\lambda=0.2$ ms. 
 
Figure~\ref{fig: strucutralconflict} shows that the measured response time is higher. The reason for this is that our 
implementation is written in Java, and the loop used to send $k$ $POSSIBLE$ 
messages by the process $P_{k+1}$ leads to computational overhead.
More precisely, when $P_{k+1}$ enters the loop to send $k$ $POSSIBLE$ messages to the different peers, 
the process $P_i$ which will get the first message sent, will set the interaction $i$ to ready and 
send back a $COMMIT$. However, $P_{k+1}$ will not treat this message before the termination of this loop.
As the actual communication time is low, the $possibleSet$ of $P_{k+1}$ may contain many  interactions, which 
increases the  {\em selection-time} (only one interaction is committed, all others must be refused).

%% file: cycleExampleN.pdf_t
\begin{picture}(0,0)%
\includegraphics{cycleExampleN.pdf}%
\end{picture}%
\setlength{\unitlength}{2279sp}%
\begingroup\makeatletter\ifx\SetFigFontNFSS\undefined%
\gdef\SetFigFontNFSS#1#2#3#4#5{%
  \reset@font\fontsize{#1}{#2pt}%
  \fontfamily{#3}\fontseries{#4}\fontshape{#5}%
  \selectfont}%
\fi\endgroup%
\begin{picture}(9179,2229)(2139,-4342)
\put(4817,-2797){\makebox(0,0)[lb]{\smash{{\SetFigFontNFSS{10}{12.0}{\rmdefault}{\bfdefault}{\updefault}{\color[rgb]{0,0,0}$a_1$}%
}}}}
\put(2656,-2491){\makebox(0,0)[lb]{\smash{{\SetFigFontNFSS{12}{14.4}{\rmdefault}{\bfdefault}{\updefault}{\color[rgb]{0,0,0}P$_1$}%
}}}}
\put(5626,-2491){\makebox(0,0)[lb]{\smash{{\SetFigFontNFSS{12}{14.4}{\rmdefault}{\bfdefault}{\updefault}{\color[rgb]{0,0,0}P$_2$}%
}}}}
\put(3781,-2716){\makebox(0,0)[lb]{\smash{{\SetFigFontNFSS{9}{10.8}{\rmdefault}{\bfdefault}{\updefault}{\color[rgb]{0,0,0}$a_1$}%
}}}}
\put(9136,-2491){\makebox(0,0)[lb]{\smash{{\SetFigFontNFSS{12}{14.4}{\rmdefault}{\bfdefault}{\updefault}{\color[rgb]{0,0,0}P$_n$}%
}}}}
\put(3376,-3211){\makebox(0,0)[lb]{\smash{{\SetFigFontNFSS{9}{10.8}{\rmdefault}{\mddefault}{\updefault}{\color[rgb]{0,0,0}S$_1$}%
}}}}
\put(6751,-2671){\makebox(0,0)[lb]{\smash{{\SetFigFontNFSS{9}{10.8}{\rmdefault}{\bfdefault}{\updefault}{\color[rgb]{0,0,0}$a_2$}%
}}}}
\put(6346,-3166){\makebox(0,0)[lb]{\smash{{\SetFigFontNFSS{9}{10.8}{\rmdefault}{\mddefault}{\updefault}{\color[rgb]{0,0,0}S$_2$}%
}}}}
\put(9856,-3166){\makebox(0,0)[lb]{\smash{{\SetFigFontNFSS{9}{10.8}{\rmdefault}{\mddefault}{\updefault}{\color[rgb]{0,0,0}S$_n$}%
}}}}
\put(10396,-2626){\makebox(0,0)[lb]{\smash{{\SetFigFontNFSS{9}{10.8}{\rmdefault}{\bfdefault}{\updefault}{\color[rgb]{0,0,0}$a_n$}%
}}}}
\put(5986,-4246){\makebox(0,0)[lb]{\smash{{\SetFigFontNFSS{10}{12.0}{\rmdefault}{\bfdefault}{\updefault}{\color[rgb]{0,0,0}$a_n$}%
}}}}
\put(2746,-2851){\makebox(0,0)[lb]{\smash{{\SetFigFontNFSS{9}{10.8}{\rmdefault}{\bfdefault}{\updefault}{\color[rgb]{0,0,0}$a_n$}%
}}}}
\put(5671,-2806){\makebox(0,0)[lb]{\smash{{\SetFigFontNFSS{9}{10.8}{\rmdefault}{\bfdefault}{\updefault}{\color[rgb]{0,0,0}$a_1$}%
}}}}
\put(9226,-2806){\makebox(0,0)[lb]{\smash{{\SetFigFontNFSS{9}{10.8}{\rmdefault}{\bfdefault}{\updefault}{\color[rgb]{0,0,0}$a_{n-1}$}%
}}}}
\end{picture}%

%% file: Tsystem.pdf_t
\begin{picture}(0,0)%
\includegraphics{Tsystem.pdf}%
\end{picture}%
\setlength{\unitlength}{2279sp}%
\begingroup\makeatletter\ifx\SetFigFontNFSS\undefined%
\gdef\SetFigFontNFSS#1#2#3#4#5{%
  \reset@font\fontsize{#1}{#2pt}%
  \fontfamily{#3}\fontseries{#4}\fontshape{#5}%
  \selectfont}%
\fi\endgroup%
\begin{picture}(6978,3804)(2284,-5023)
\put(2835,-4574){\makebox(0,0)[lb]{\smash{{\SetFigFontNFSS{8}{9.6}{\rmdefault}{\mddefault}{\updefault}{\color[rgb]{0,0,0}S$_1$}%
}}}}
\put(2341,-4111){\makebox(0,0)[lb]{\smash{{\SetFigFontNFSS{11}{13.2}{\rmdefault}{\bfdefault}{\updefault}{\color[rgb]{0,0,0}P$_1$}%
}}}}
\put(3511,-4156){\makebox(0,0)[lb]{\smash{{\SetFigFontNFSS{8}{9.6}{\rmdefault}{\bfdefault}{\updefault}{\color[rgb]{0,0,0}$a_1$}%
}}}}
\put(5265,-4574){\makebox(0,0)[lb]{\smash{{\SetFigFontNFSS{8}{9.6}{\rmdefault}{\mddefault}{\updefault}{\color[rgb]{0,0,0}S$_2$}%
}}}}
\put(4771,-4111){\makebox(0,0)[lb]{\smash{{\SetFigFontNFSS{11}{13.2}{\rmdefault}{\bfdefault}{\updefault}{\color[rgb]{0,0,0}P$_2$}%
}}}}
\put(5941,-4156){\makebox(0,0)[lb]{\smash{{\SetFigFontNFSS{8}{9.6}{\rmdefault}{\bfdefault}{\updefault}{\color[rgb]{0,0,0}$a_2$}%
}}}}
\put(8055,-4574){\makebox(0,0)[lb]{\smash{{\SetFigFontNFSS{8}{9.6}{\rmdefault}{\mddefault}{\updefault}{\color[rgb]{0,0,0}S$_k$}%
}}}}
\put(7561,-4111){\makebox(0,0)[lb]{\smash{{\SetFigFontNFSS{11}{13.2}{\rmdefault}{\bfdefault}{\updefault}{\color[rgb]{0,0,0}P$_k$}%
}}}}
\put(8731,-4156){\makebox(0,0)[lb]{\smash{{\SetFigFontNFSS{8}{9.6}{\rmdefault}{\bfdefault}{\updefault}{\color[rgb]{0,0,0}$a_k$}%
}}}}
\put(4006,-3481){\makebox(0,0)[lb]{\smash{{\SetFigFontNFSS{9}{10.8}{\rmdefault}{\bfdefault}{\updefault}{\color[rgb]{0,0,0}$a_1$}%
}}}}
\put(5761,-3481){\makebox(0,0)[lb]{\smash{{\SetFigFontNFSS{9}{10.8}{\rmdefault}{\bfdefault}{\updefault}{\color[rgb]{0,0,0}$a_2$}%
}}}}
\put(5626,-1636){\makebox(0,0)[lb]{\smash{{\SetFigFontNFSS{9}{10.8}{\rmdefault}{\bfdefault}{\updefault}{\color[rgb]{0,0,0}$a_1$}%
}}}}
\put(4321,-1591){\makebox(0,0)[lb]{\smash{{\SetFigFontNFSS{12}{14.4}{\rmdefault}{\bfdefault}{\updefault}{\color[rgb]{0,0,0}P$_{k+1}$}%
}}}}
\put(4816,-2536){\makebox(0,0)[lb]{\smash{{\SetFigFontNFSS{9}{10.8}{\rmdefault}{\bfdefault}{\updefault}{\color[rgb]{0,0,0}$a_2$}%
}}}}
\put(6526,-1861){\makebox(0,0)[lb]{\smash{{\SetFigFontNFSS{9}{10.8}{\rmdefault}{\bfdefault}{\updefault}{\color[rgb]{0,0,0}$a_k$}%
}}}}
\put(5401,-2131){\makebox(0,0)[lb]{\smash{{\SetFigFontNFSS{9}{10.8}{\rmdefault}{\mddefault}{\updefault}{\color[rgb]{0,0,0}S$_{k+1}$}%
}}}}
\put(8101,-3481){\makebox(0,0)[lb]{\smash{{\SetFigFontNFSS{9}{10.8}{\rmdefault}{\bfdefault}{\updefault}{\color[rgb]{0,0,0}$a_k$}%
}}}}
\end{picture}%

%% file: diningPhilo.tex
\vspace{-\baselineskip}
\subsection{The dining philosophers example}
\vspace{-0.5\baselineskip}
We have carried out a series of tests on the well-known Dining philosophers problem.
We consider a variant of the dining philosophers problem inspired from
\cite{Padovani08} and we propose to deal with this problem using priorities.
Philosophers are seen as processes who
provide thoughts if they are given two forks. These forks represent a shared resource. 
 A problem may arise if each philosopher grabs the fork on its right, and then 
waits for the fork on its left to be released. In this case a deadlock occurs and all philosophers
 starve.
\begin{center}
 \begin{figure}
    \begin{center}
  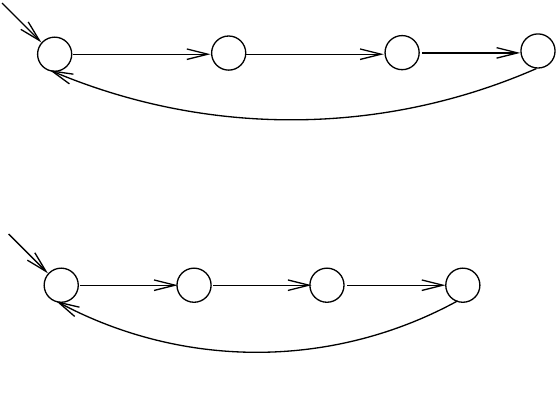
      \hspace{1cm}
      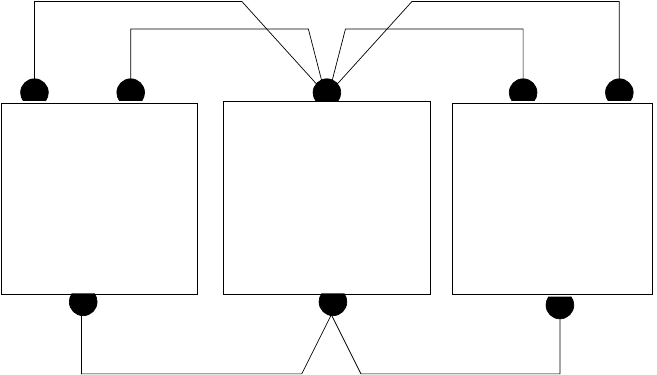
   \end{center}    
     \caption{The dining philosophers problem with priorities.}\label{fig:philoPb}
  \end{figure}
\vspace{-0.5\baselineskip}
\end{center}

This deadlock can be avoided by giving higher priority to  requests
 closer to completion. 
The priority order that is needed here is $\{\mathit{fork}_1^\alpha <
\mathit{fork}_2^\beta,\ \mathit{fork}_1^\beta < \mathit{fork}_2^\alpha
\}$. For readability reasons, in Figure~\ref{fig:philoPb}, the interaction $\mathit{fork}_{1,2}^{\alpha,\beta}$ 
in the behavior of the process $Forks$ corresponds to the interactions $\{\mathit{fork}_1^\alpha,\ \mathit{fork}_2^\alpha,\ 
\mathit{fork}_1^\beta,\ \mathit{fork}_2^\beta\}$ of the two philosophers.
As the process $Forks$ participates in all interactions involved in these priorities, 
$Forks$ is designated  negotiator for involved interactions and can ensure locally
that priorities are respected.
Experiments have been carried out for the system with the mentioned priorities 
(depicted in Figure~\ref{fig:philoPb}); then we have also considered
 a system with two philosophers and separate processes for each fork, where deadlock is avoided
 by the fact that both philosophers first request $Fork_1$, and  then $Fork_2$. 

\vspace{-0.5\baselineskip}
\newcolumntype{M}[1]{>{\raggedright}m{#1}}
\begin{table}[h]
\centering 
\begin{tabular}{|l | M{2.5cm}| M{3.5cm}| M{4.5cm}|} %
 \hline
Dining philosophers & {\em Message-count} & {\em Execution-time}(ms)  & {\em Execution-time$_{Philo^*}$}(ms)\tabularnewline
\hline 
With priorities & 6  & 8   & 30\tabularnewline
 \hline 
Without priorities & 6  & 11  & 45\tabularnewline
\hline 
\end{tabular}
\caption{Message count for the dining philosophers}
\label{table:msgcount}
\end{table} 
\vspace{-0.5\baselineskip}
Table~\ref{table:msgcount} shows our measurements for the message-count and the response-time metrics 
for both systems.  We have also measured the average
time required for one philosopher to execute a complete cycle (take forks, think and release forks) which we denote by {\em Execution-Time$_{Philo^*}$}.\smallskip

We observe that the number of messages exchanged is identical in these two systems. Indeed, priorities 
 are local 
thus do not induce additional messages. However, using priorities leads to a slight increase of
the  {\em execution-time}, as we have already observed in our first example, and the explanation
remains unchanged. 
An additional reason is that the system with priorities has only one  process to handle both forks, 
this making the system less concurrent than the system without priorities. This effect of concurrency
 is particularly visible  in the results for 
{\em Execution-Time$_{Philo^*}$}.

%% file: philosopherEx.pdf_t
\begin{picture}(0,0)%
\includegraphics{philosopherEx.pdf}%
\end{picture}%
\setlength{\unitlength}{1657sp}%
\begingroup\makeatletter\ifx\SetFigFontNFSS\undefined%
\gdef\SetFigFontNFSS#1#2#3#4#5{%
  \reset@font\fontsize{#1}{#2pt}%
  \fontfamily{#3}\fontseries{#4}\fontshape{#5}%
  \selectfont}%
\fi\endgroup%
\begin{picture}(6357,4711)(1074,-3325)
\put(1861,914){\makebox(0,0)[lb]{\smash{{\SetFigFontNFSS{8}{9.6}{\familydefault}{\mddefault}{\updefault}{\color[rgb]{0,0,0}\normalsize{$\mathit{fork}_{1,2}^{\alpha, \beta}$}}%
}}}}
\put(2986,-526){\makebox(0,0)[lb]{\smash{{\SetFigFontNFSS{8}{9.6}{\familydefault}{\mddefault}{\updefault}{\color[rgb]{0,0,0}\normalsize{The behavior of $Forks$}}%
}}}}
\put(3511,-1726){\makebox(0,0)[lb]{\smash{{\SetFigFontNFSS{8}{9.6}{\familydefault}{\mddefault}{\updefault}{\color[rgb]{0,0,0}\normalsize{$\mathit{fork}_{2}^*$}}%
}}}}
\put(3916,884){\makebox(0,0)[lb]{\smash{{\SetFigFontNFSS{8}{9.6}{\familydefault}{\mddefault}{\updefault}{\color[rgb]{0,0,0}\normalsize{$\mathit{fork}_{1,2}^{\alpha, \beta}$}}%
}}}}
\put(5941,884){\makebox(0,0)[lb]{\smash{{\SetFigFontNFSS{8}{9.6}{\familydefault}{\mddefault}{\updefault}{\color[rgb]{0,0,0}\normalsize{$\mathit{return}$}}%
}}}}
\put(3601,164){\makebox(0,0)[lb]{\smash{{\SetFigFontNFSS{8}{9.6}{\familydefault}{\mddefault}{\updefault}{\color[rgb]{0,0,0}\normalsize{$\mathit{return}$}}%
}}}}
\put(3331,-2491){\makebox(0,0)[lb]{\smash{{\SetFigFontNFSS{8}{9.6}{\familydefault}{\mddefault}{\updefault}{\color[rgb]{0,0,0}\normalsize{$\mathit{return}^*$}}%
}}}}
\put(2971,-3211){\makebox(0,0)[lb]{\smash{{\SetFigFontNFSS{8}{9.6}{\familydefault}{\mddefault}{\updefault}{\color[rgb]{0,0,0}\normalsize{The behavior of $Philo_{*}$}}%
}}}}
\put(5086,-1726){\makebox(0,0)[lb]{\smash{{\SetFigFontNFSS{8}{9.6}{\familydefault}{\mddefault}{\updefault}{\color[rgb]{0,0,0}\normalsize{$\mathit{return}^*$}}%
}}}}
\put(2071,-1726){\makebox(0,0)[lb]{\smash{{\SetFigFontNFSS{8}{9.6}{\familydefault}{\mddefault}{\updefault}{\color[rgb]{0,0,0}\normalsize{$\mathit{fork}_1^*$}}%
}}}}
\end{picture}%

%% file: philoStruct.pdf_t
\begin{picture}(0,0)%
\includegraphics{philoStruct.pdf}%
\end{picture}%
\setlength{\unitlength}{1782sp}%
\begingroup\makeatletter\ifx\SetFigFontNFSS\undefined%
\gdef\SetFigFontNFSS#1#2#3#4#5{%
  \reset@font\fontsize{#1}{#2pt}%
  \fontfamily{#3}\fontseries{#4}\fontshape{#5}%
  \selectfont}%
\fi\endgroup%
\begin{picture}(6950,3984)(-1811,-5743)
\put(4222,-3359){\makebox(0,0)[lb]{\smash{{\SetFigFontNFSS{8}{9.6}{\rmdefault}{\mddefault}{\updefault}{\color[rgb]{0,0,0}$fork_2^{\beta}$}%
}}}}
\put(3317,-3359){\makebox(0,0)[lb]{\smash{{\SetFigFontNFSS{8}{9.6}{\rmdefault}{\mddefault}{\updefault}{\color[rgb]{0,0,0}$\mathit{fork_1^{\beta}}$}%
}}}}
\put(-1760,-3300){\makebox(0,0)[lb]{\smash{{\SetFigFontNFSS{8}{9.6}{\rmdefault}{\mddefault}{\updefault}{\color[rgb]{0,0,0}$fork_1^{\alpha}$}%
}}}}
\put(-815,-3300){\makebox(0,0)[lb]{\smash{{\SetFigFontNFSS{8}{9.6}{\rmdefault}{\mddefault}{\updefault}{\color[rgb]{0,0,0}$fork_2^{\alpha}$}%
}}}}
\put(1231,-4182){\makebox(0,0)[lb]{\smash{{\SetFigFontNFSS{8}{9.6}{\rmdefault}{\mddefault}{\updefault}{\color[rgb]{0,0,0}$\mathbf{Forks}$}%
}}}}
\put(-1209,-4182){\makebox(0,0)[lb]{\smash{{\SetFigFontNFSS{8}{9.6}{\rmdefault}{\mddefault}{\updefault}{\color[rgb]{0,0,0}$\normalsize{\mathbf{Philo_\alpha}}$}%
}}}}
\put(3592,-4182){\makebox(0,0)[lb]{\smash{{\SetFigFontNFSS{8}{9.6}{\rmdefault}{\mddefault}{\updefault}{\color[rgb]{0,0,0}$\normalsize{\mathbf{Philo_\beta}}$}%
}}}}
\put(1271,-3359){\makebox(0,0)[lb]{\smash{{\SetFigFontNFSS{8}{9.6}{\rmdefault}{\mddefault}{\updefault}{\color[rgb]{0,0,0}$\mathit{fork_{1,2}^{\alpha, \beta}}$}%
}}}}
\put(-1304,-4786){\makebox(0,0)[lb]{\smash{{\SetFigFontNFSS{8}{9.6}{\rmdefault}{\mddefault}{\updefault}{\color[rgb]{0,0,0}$\mathit{return}$}%
}}}}
\put(3871,-4786){\makebox(0,0)[lb]{\smash{{\SetFigFontNFSS{8}{9.6}{\rmdefault}{\mddefault}{\updefault}{\color[rgb]{0,0,0}$\mathit{return}$}%
}}}}
\put(1396,-4741){\makebox(0,0)[lb]{\smash{{\SetFigFontNFSS{8}{9.6}{\rmdefault}{\mddefault}{\updefault}{\color[rgb]{0,0,0}$\mathit{return}$}%
}}}}
\end{picture}%

%% file: conclusion.tex
In this paper, we have presented and evaluated an implementation of the algorithm proposed in \cite{TRJLAP}, which 
defines a transformation of a global specification of a  component-based system with priorities into 
a distributed system, in which every component becomes a process that may be executed on a different 
physical machine, and for this purpose is composed with a local controller exchanging messages
with peer controllers.
We analyze the performance of the algorithm on hand of a number of experiments and measure 3 different metrics
by executing the implementation for different systems.
These results show that our algorithm behaves as expected.
A comparison with the $\alpha$-core algorithm is performed based on results available in the literature,
which shows that our algorithm requires a much smaller amount of messages for systems without priorities.
$\alpha$-core does not handle priorities, and, for the time being, our algorithm does not handle multi-party
synchronization which $\alpha$-core does.

More experimentation is required, and different improvements of our algorithm are envisaged.
In particular, beyond the adaptation to multi-party interactions, we plan adapting knowledge-based methods as proposed  
\cite{QuintonGP10, BasuBPS09}. This adaptation is not straightforward, as these algorithms are applied
at the level of synchronizing processes, thus ignoring how the synchronization are realized in terms of 
message exchanges, whereas our algorithm addresses exactly this lower level, where additional knowledge
may be exploited to decrease the number of messages exchanged without significantly increasing the local memory or the complexity of the local algorithms.

%% file: algorithms.tex
\vspace{-2\baselineskip}
{\small
\begin{algorithm}\caption{Main: \hspace{1cm} \textbf{Input:} $possibleSet \neq \emptyset$ \hspace{0.5cm}
\textbf{Output:}  interaction $a$
}
\label{Main}
\begin{algorithmic}[]
\REQUIRE $toNegotiate = \{a\in possibleSet\ |\ negotiator(a) = P \}$\\
\textbf{Input:} set of interactions $possibleSet \neq \emptyset$ \hspace{0.5cm}
\textbf{Output:}  interaction $i$
\STATE $prioFree = \{a\in possibleSet\ |\ \not\exists b\,.\, b<a\}$
\STATE $waitingSet \longleftarrow \emptyset$ 
\STATE \textbf{checking global readiness:} 
\STATE $notReadySet \longleftarrow \emptyset$ 
\STATE $readySet \longleftarrow \emptyset$
\STATE $lessPrio(a)= \{b\in readySet |\ b < a \}\}$
\FORALL{$a\in possibleSet$}
\STATE \textbf{send} $\POSSIBLE(a)$
\ENDFOR
\STATE \textbf{create} WaitingForCommit($possibleSet$)
\IF{\textbf{receive} $\POSSIBLE(a)$ and $a\in toNegotiate$}
\STATE \textbf{create} Negotiate($a$) and $readySet \longleftarrow readySet \cup \{a\}$ and 
\FORALL{$b\in lessPrio(a)$}
\STATE \textbf{kill} Negotiate($b$) 
\ENDFOR
\ENDIF
\STATE \textbf{WHEN} $\exists \ a \ s.t.\ $ Negotiate($a$)= $OK$ or (\textbf{receive} $\POSSIBLE(a)$ and $a \in prioFree$)
\STATE \textbf{call} TryToCommit($a$) and \textbf{kill} WaitingForCommit($possibleSet$) and $\forall b \in readySet$ \textbf{kill} Negotiate($b$)
\IF{TryToCommit($a$)= $OK$}
\STATE \textbf{return} $a$
\ELSE 
\STATE \textbf{goto checking global readiness} 
\ENDIF
\IF{$\forall a \in readySet$ Negotiate($a$)= $NOK$ }
\STATE \textbf{goto checking global readiness} 
\ENDIF
\IF{\textbf{receive} $\REFUSE(b)$ and $b \in readySet$}
\STATE \textbf{kill} Negotiate($b$) and $readySet \longleftarrow readySet \backslash \{b\}$
\ENDIF
\IF{\textbf{receive} $\POSSIBLE(b)$ and $b \in possibleSet\backslash\{toNegotiate\cup prioFree\}$}
\STATE \textbf{send}  $\POSSIBLE(b)$ and $readySet \longleftarrow readySet \cup \{b\}$
\ENDIF
\IF{\textbf{receive} $\NOTPOSSIBLE(b)$ and $b \in possibleSet\backslash prioFree$}
\STATE $notReadySet \longleftarrow notReadySet \cup \{b\}$
\ENDIF
\IF{\textbf{receive} $\POSSIBLE(b)$ and $b \not \in possibleSet$}
\STATE \textbf{send}  $\NOTPOSSIBLE(b)$
\ENDIF
\end{algorithmic}
\end{algorithm}
}
\vspace{-\baselineskip}
\begin{algorithm}\caption{Negotiate: \hspace{1cm} \textbf{Input:} interaction $a$\hspace{0.5cm}\textbf{Output:} $OK$ or $NOK$}
\label{negotiate}
\begin{algorithmic}[]
\REQUIRE $higherPrio(a) = \{c\ |\ a<c\}$
 \FORALL{$b \in higherPrio(a)$}
 \STATE \textbf{send} $\READY(b)$
 \ENDFOR
\WHILE{$higherPrio(a) \neq \emptyset$}
\IF{\textbf{receive} $\READY(b)$}
\STATE \textbf{return} $NOK$
\ELSIF{\textbf{receive} $\NOTREADY(b)$}
\STATE $higherPrio(a) \longleftarrow higherPrio(a) \backslash \{b\}$
\ENDIF
\ENDWHILE
  \STATE \textbf{return} $OK$ 
\end{algorithmic}
\end{algorithm}
\vspace{-\baselineskip}
\begin{algorithm}\caption{WaitingForCommit: \hspace{1cm} \textbf{Input:} $possibleSet$\hspace{0.5cm} \textbf{Output:} interaction $a$}
  \label{waitingForCommit}
  \begin{algorithmic}[]
  \REQUIRE 
    \IF{$waitingSet \neq \emptyset$}
    \STATE \textbf{choose} $a\in waitingSet$ and \textbf{kill} main and \textbf{send} $\COMMIT(a)$ and
     \textbf{send} $\REFUSE(b)$ for all $b$ in $possibleSet$ and \textbf{goto} $Busy(a)$
    \ELSIF{$waitingSet = \emptyset$ and \textbf{receive}$\COMMIT(a)$ and $a \in possibleSet \backslash toNegotiate$} 
     \STATE \textbf{kill} main and \textbf{send} $\COMMIT(a)$ and
     \textbf{send} $\REFUSE(b)$ for all $b$ in $possibleSet$ and \textbf{goto} $Busy(a)$
    \ENDIF
  \IF{\textbf{receive} $\COMMIT(a)$ and $a \not \in possibleSet$}
  \STATE \textbf{send}  $\REFUSE(a)$
  \ENDIF
  \end{algorithmic}
\end{algorithm}
\vspace{-\baselineskip}
 \begin{algorithm}\caption{TryToCommit: \hspace{1cm} \textbf{Input:} set of interactions $readySet$ \hspace{0.5cm}
    \textbf{Output:} $OK$ or $NOK$}
  \label{tryToCommit}
  \begin{algorithmic}[]
    \REQUIRE 
    \STATE \textbf{send} $\COMMIT(a)$
    \IF{\textbf{receive} $\COMMIT(a)$}
    \STATE \textbf{return} $OK$ and \textbf{send} $\forall b \in$ readySet$\setminus\{a\}$ $\REFUSE(b)$  
    \ELSIF{\textbf{receive} $\COMMIT(b)$ and $b \neq a$ and ($b \not \in cycle(a)$ or( $b \in cycle(a) \wedge P_b = Cyclebreaker$)) }
    \STATE $waitingSet \longleftarrow waitingSet \cup \{b\}$
    \ELSIF{\textbf{receive} $\COMMIT(b)$ and $b \neq a$ and $b \in cycle(a)$ and $P_b \neq Cyclebreaker$}
    \STATE \textbf{send} $\REFUSE(b)$ and $readySet \longleftarrow readySet \backslash \{b\}$
    \ELSIF{\textbf{receive} $\REFUSE(a)$} 
    \STATE \textbf{return} $NOK$
    \ENDIF
  \end{algorithmic}
\end{algorithm}

%% file: main.bbl
\begin{thebibliography}{10}
\providecommand{\bibitemstart}[1]{\bibitem{#1}}
\providecommand{\bibitemend}{}
\providecommand{\bibliographystart}{}
\providecommand{\bibliographyend}{}
\providecommand{\url}[1]{\texttt{#1}}
\providecommand{\urlprefix}{Available at }
\providecommand{\bibinfo}[2]{#2}
\bibliographystart

\bibitemstart{Bagrodia89}
\bibinfo{author}{Rajive Bagrodia} (\bibinfo{year}{1989}):
  \emph{\bibinfo{title}{Synchronization of Asynchronous Processes in {CSP}}}.
\newblock {\sl \bibinfo{journal}{ACM Trans. Program. Lang. Syst.}}
  \bibinfo{volume}{11}(\bibinfo{number}{4}), pp. \bibinfo{pages}{585--597}.
\newblock \urlprefix\url{http://doi.acm.org/10.1145/69558.69561}.
\bibitemend

\bibitemstart{BasuBPS09}
\bibinfo{author}{Ananda Basu}, \bibinfo{author}{Saddek Bensalem},
  \bibinfo{author}{Doron Peled} \& \bibinfo{author}{Joseph Sifakis}
  (\bibinfo{year}{2009}): \emph{\bibinfo{title}{Priority Scheduling of
  Distributed Systems Based on Model Checking}}.
\newblock In: {\sl \bibinfo{booktitle}{CAV}}, {\sl \bibinfo{series}{Lecture
  Notes in Computer Science}} \bibinfo{volume}{5643},
  \bibinfo{publisher}{Springer}, pp. \bibinfo{pages}{79--93}.
\newblock \urlprefix\url{http://dx.doi.org/10.1007/978-3-642-02658-4_10}.
\bibitemend

\bibitemstart{BasuBS06}
\bibinfo{author}{Ananda Basu}, \bibinfo{author}{Marius Bozga} \&
  \bibinfo{author}{Joseph Sifakis} (\bibinfo{year}{2006}):
  \emph{\bibinfo{title}{Modeling Heterogeneous Real-time Components in {BIP}}}.
\newblock In: {\sl \bibinfo{booktitle}{Proc. of SEFM'06}},
  \bibinfo{publisher}{IEEE Computer Society}, pp. \bibinfo{pages}{3--12}.
\newblock
  \urlprefix\url{http://doi.ieeecomputersociety.org/10.1109/SEFM.2006.27}.
\bibitemend

\bibitemstart{TRJLAP}
\bibinfo{author}{Imene Ben-Hafaiedh}, \bibinfo{author}{Susanne Graf} \&
  \bibinfo{author}{Sophie Quinton} (\bibinfo{year}{2010}):
  \emph{\bibinfo{title}{Building Distributed Controllers for Systems with
  Priorities}}.
\newblock
  \urlprefix\url{http://www-verimag.imag.fr/Technical-Reports,264.html?lang=en%
&number=TR-2010-15}.
\bibitemend

\bibitemstart{BliudzeS07}
\bibinfo{author}{Simon Bliudze} \& \bibinfo{author}{Joseph Sifakis}
  (\bibinfo{year}{2007}): \emph{\bibinfo{title}{The algebra of connectors:
  structuring interaction in {BIP}}}.
\newblock In: {\sl \bibinfo{booktitle}{Proc. of EMSOFT'07}},
  \bibinfo{publisher}{ACM Press}, pp. \bibinfo{pages}{11--20}.
\newblock \urlprefix\url{http://doi.acm.org/10.1145/1289927.1289935}.
\bibitemend

\bibitemstart{ChandyLamport85}
\bibinfo{author}{K.~Mani Chandy} \& \bibinfo{author}{Leslie Lamport}
  (\bibinfo{year}{1985}): \emph{\bibinfo{title}{Distributed snapshots:
  determining global states of distributed systems}}.
\newblock {\sl \bibinfo{journal}{ACM Trans. Comput. Syst.}}
  \bibinfo{volume}{3}(\bibinfo{number}{1}), pp. \bibinfo{pages}{63--75}.
\bibitemend

\bibitemstart{GoesslerS05}
\bibinfo{author}{Gregor G{\"o}{\ss}ler} \& \bibinfo{author}{Joseph Sifakis}
  (\bibinfo{year}{2005}): \emph{\bibinfo{title}{Composition for component-based
  modeling}}.
\newblock {\sl \bibinfo{journal}{Sci. Comput. Program.}}
  \bibinfo{volume}{55}(\bibinfo{number}{1-3}), pp. \bibinfo{pages}{161--183}.
\newblock \urlprefix\url{http://dx.doi.org/10.1016/j.scico.2004.05.014}.
\bibitemend

\bibitemstart{MPIPortable}
\bibinfo{author}{William Gropp}, \bibinfo{author}{Ewing Lusk} \&
  \bibinfo{author}{Anthony Skjellum} (\bibinfo{year}{1999}):
  \emph{\bibinfo{title}{Using MPI (2nd ed.): portable parallel programming with
  the message-passing interface}}.
\newblock \bibinfo{publisher}{MIT Press}, \bibinfo{address}{Cambridge, MA,
  USA}.
\bibitemend

\bibitemstart{Padovani08}
\bibinfo{author}{Luca Padovani} (\bibinfo{year}{2008}):
  \emph{\bibinfo{title}{Contract-Directed Synthesis of Simple Orchestrators}}.
\newblock In: {\sl \bibinfo{booktitle}{Proc. of CONCUR'08}}, {\sl
  \bibinfo{series}{LNCS}} \bibinfo{volume}{5201}, pp.
  \bibinfo{pages}{131--146}.
\newblock \urlprefix\url{http://dx.doi.org/10.1007/978-3-540-85361-9_13}.
\bibitemend

\bibitemstart{PerezCT04}
\bibinfo{author}{Jos{\'e}~Antonio P{\'e}rez}, \bibinfo{author}{Rafael
  Corchuelo} \& \bibinfo{author}{Miguel Toro} (\bibinfo{year}{2004}):
  \emph{\bibinfo{title}{An order-based algorithm for implementing multiparty
  synchronization}}.
\newblock {\sl \bibinfo{journal}{Concurrency - Practice and Experience}}
  \bibinfo{volume}{16}(\bibinfo{number}{12}), pp. \bibinfo{pages}{1173--1206}.
\newblock \urlprefix\url{http://dx.doi.org/10.1002/cpe.903}.
\bibitemend

\bibitemstart{SnirOttoHussLedermanEtAl96}
\bibinfo{author}{M.~Snir}, \bibinfo{author}{S.~W. Otto},
  \bibinfo{author}{S.~Huss-Lederman}, \bibinfo{author}{D.~W. Walker} \&
  \bibinfo{author}{J~Dongarra} (\bibinfo{year}{1996}):
  \emph{\bibinfo{title}{MPI: The complete reference}}.
\newblock \bibinfo{publisher}{MIT Press}, \bibinfo{address}{Cambridge, MA}.
\bibitemend

\bibitemstart{QuintonGP10}
\bibinfo{author}{Doron~Peled Susanne~Graf} \& \bibinfo{author}{Sophie Quinton}
  (\bibinfo{year}{2010}): \emph{\bibinfo{title}{Achieving Distributed Control
  Through Model Checking}}.
\newblock In: {\sl \bibinfo{booktitle}{CAV}}.
\bibitemend

\bibliographyend
\end{thebibliography}
